\documentclass[]{interact}

\usepackage{epstopdf}
\usepackage{subfigure}

\usepackage{natbib}
\bibpunct[, ]{(}{)}{;}{a}{}{,}
\renewcommand\bibfont{\fontsize{10}{12}\selectfont}

\theoremstyle{plain}

\theoremstyle{definition}

\theoremstyle{remark}

\usepackage{bm}
\usepackage{multirow} 
\usepackage{makecell}
\usepackage{scalerel}
\usepackage{subfigure}
\usepackage{color}

\usepackage{natbib}
\bibpunct[, ]{(}{)}{;}{a}{}{,}
\renewcommand\bibfont{\fontsize{10}{12}\selectfont}

\begin{document}

\articletype{ARTICLE TEMPLATE}

\title{Topology Optimization for Large-Scale Additive Manufacturing: Generating designs tailored to the deposition nozzle size}

\author{
\name{E.~Fern\'{a}ndez\textsuperscript{a}\thanks{Eduardo Fern\'{a}ndez, orcid: 0000-0003-3735-1886, Email: efsanchez@uliege.be}, C.~Ayas\textsuperscript{b}\thanks{Can Ayas, orcid: 0000-0002-9472-7424,  Email: C.Ayas@tudelft.nl}, M.~Langelaar\textsuperscript{b}\thanks{Matthijs Langelaar, orcid: 0000-0003-2106-2246, Email: M.Langelaar@tudelft.nl} and P.~Duysinx\textsuperscript{a}\thanks{Pierre Duysinx, orcid: 0000-0001-7870-3628, Email: P.Duysinx@uliege.be}}
\affil{\textsuperscript{a}Department of Aerospace and Mechanical Engineering, University of Liege, All\'{e}e de la D\'{e}couverte 13A, B52, 4000, Liege, Belgium; \textsuperscript{b}Faculty of Mechanical, Maritime and Materials Engineering, Delft University of Technology, Mekelweg 2, 2628 CD Delft, The Netherlands.}
}

\maketitle

\begin{abstract}
Additive Manufacturing (AM) processes intended for large scale components deposit large volumes of material to shorten process duration. This reduces the resolution of the AM process, which is typically defined by the size of the deposition nozzle. If the resolution limitation is not considered when designing for Large-Scale Additive Manufacturing (LSAM), difficulties can arise in the manufacturing process, which may require the adaptation of the deposition parameters. This work incorporates the nozzle size constraint into Topology Optimization (TO) in order to generate optimized designs suitable to the process resolution. This article proposes and compares two methods, which are based on existing TO techniques that enable control of minimum and maximum member size, and of minimum cavity size. The first method requires the minimum and maximum member size to be equal to the deposition nozzle size, thus design features of uniform width are obtained in the optimized design. The second method defines the size of the solid members sufficiently small for the resulting structure to resemble a structural skeleton, which can be interpreted as the deposition path. Through filtering and projection techniques, the thin structures are thickened according to the chosen nozzle size. Thus, a topology tailored to the size of the deposition nozzle is obtained along with a deposition proposal. The methods are demonstrated and assessed using 2D and 3D benchmark problems.
\end{abstract}


\section{Introduction}

Topology Optimization (TO) and Additive Manufacturing (AM) are recognized as promising technologies for the realization of high performance structural components due to their ability to create designs with unprecedented sophistication. The potential of these technologies has caught intense interest across a wide range of industries \citep{Zhu2016,Culmone2019,Tino2020}, which promotes their accelerated development. While AM techniques are becoming more versatile, stable and precise \citep{Li2019,Bajaj2020}, TO is adapting to the limitations of the AM processes \citep{Liu2018,Meng2019}.

Topology Optimization is well known for its ability to generate highly efficient designs. However, TO designs often come with high geometrical complexity which hinders manufacturing for even the most advanced AM processes. This drawback has established a research field seeking to facilitate the manufacturability of TO design \citep{Liu2018}. To date, a number of TO methods have been proposed for specific AM processes, such as powder-based processes, stereolithography or Fused Deposition Modeling (FDM). For instance, there are methods to prevent closed cavities in optimized designs in order to ensure the removal of support structures or unfused powder \citep{Li2016,Langelaar2019,Gaynor2020}. To reduce or avoid the use of support structures, the maximum permissible overhang angle of parts can be controlled during TO \citep{Langelaar2016,Gaynor2016,Zhang2020,van2020} along with the printing direction \citep{Langelaar2018,Wang2020}. In addition, minimum member size and minimum cavity size can be imposed in TO to ensure their printability \citep{Wang2011,Pellens2019}. More complex methods take into account thermal residual stresses in TO \citep{Allaire2018}, or incorporate continuity constraints to facilitate the printability of multiple materials \citep{Yu2020}. Experimental studies \citep{Fu2019} and industrial applications \citep{Zhu2016} have demonstrated that such AM-specific TO methods enhance the manufacturability of the optimized design when it comes to FDM, stereolithography or powder-based AM processes.

\begin{figure}
    \centering
	\includegraphics[width=0.23\linewidth]{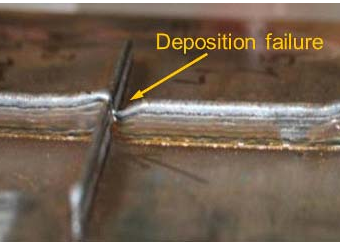}
	~
	\includegraphics[width=0.23\linewidth]{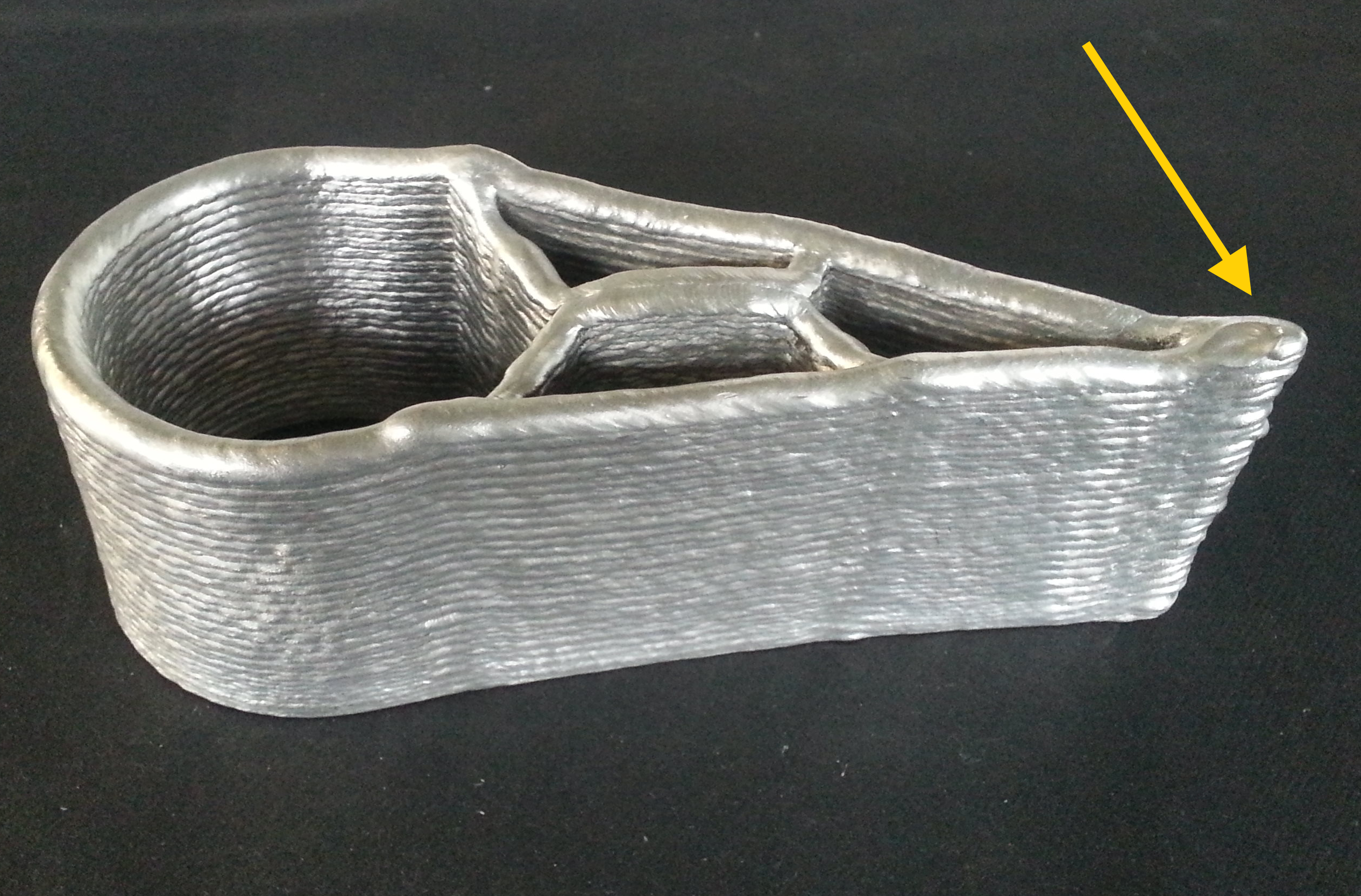}
	~
	\includegraphics[width=0.23\linewidth]{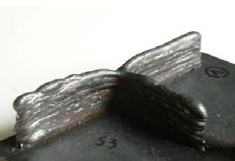}
	~	
	\includegraphics[width=0.23\linewidth]{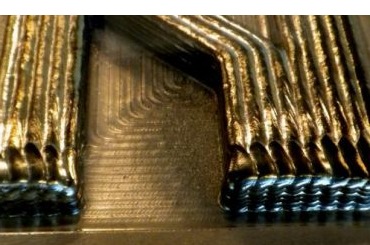}
	\\
	\subfigure[Crossover weld paths \citep{Mehnen2014}.]{
		\resizebox*{3cm}{!}{
		\includegraphics[width=0.05\linewidth]{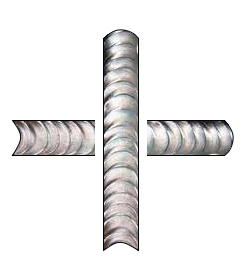}
		\label{fig:WAAM_Constraints_a}}
	}
	~
	\subfigure[Sharp corner.]{	
		\resizebox*{3cm}{!}{	
		\includegraphics[width=0.03\linewidth]{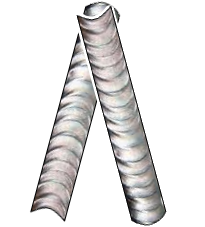} 
		\label{fig:WAAM_Constraints_c}}
	}	 
	~
	\subfigure[T-type connection \citep{Venturini2016}.]{
		\resizebox*{3cm}{!}{
		\includegraphics[width=0.05\linewidth]{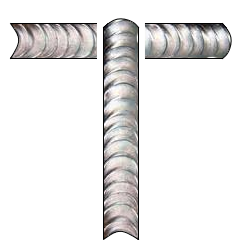}
		\label{fig:WAAM_Constraints_b}}
	}	
	~
	\subfigure[Discrete welding beads \citep{Jackson2016}.]{
		\resizebox*{3cm}{!}{		
		\includegraphics[width=0.07\linewidth]{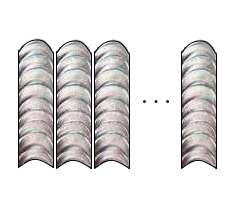}
		\label{fig:WAAM_Constraints_d}}
	}		
	
	\caption{Some geometric features to be considered when designing for a Large-Scale Additive Manufacturing (LSAM) process. Figure 1b is courtesy of Institute Soudure (www.isgroupe.com). Figure 1d is courtesy of AML3D (www.aml3d.com).}			
	\label{fig:WAAM_Constraints}
\end{figure}

There is an emerging interest in the industry for Large-Scale Additive Manufacturing (LSAM) processes, as they are able to produce components in short time spans \citep{Lim2012}, either to manufacture large-scale parts \citep{Greer2019}, or to speed up mass production \citep{Bishop2020}. Among others, processes such as Direct Metal Deposition (DMD), Wire-Arc Additive Manufacturing (WAAM), 3D Concrete Printing (3DCP), and large-scale Fused Deposition Modelling (FDM) stand out. These LSAM processes achieve high production rates by depositing relatively large volumes of material per unit time. Though they operate using the layer-by-layer principle, they exhibit distinctive limitations compared to low production rate processes, such as powder bed fusion. For example, the large amount of deposited material complicates the deposition path planning \citep{Jiang2020}, since crossover of deposition paths leads to local agglomerations of material that accumulate geometrical defects from layer to layer \citep{Mehnen2014}, as shown in Fig.~\ref{fig:WAAM_Constraints_a}. Similarly, design features with sharp corners and small radii of curvature may also cause local agglomeration of material \citep{Geng2017,Comminal2019}, as shown in Fig.~\ref{fig:WAAM_Constraints_c}, while deposition paths with T-joints can induce defects during stops and starts, such as excess or lack of deposited material \citep{Venturini2016}, as shown in Fig.~\ref{fig:WAAM_Constraints_b}. These difficulties have been included in path planning algorithms \citep{Liu2020}. However, when the design to be manufactured has features with geometrical details smaller than the nominal resolution of the LSAM process, the path planning becomes extremely difficult, since the deposition process parameters must be included in the planning algorithm. For example, the nozzle size can be reduced, but if this is not possible, then the feed rate and travel speed can be modified, or the deposition beads can be overlapped. Changing the deposition parameters can reduce the process stability, induce geometrical or metallurgical defects and demand extensive testing, thus increasing production costs \citep{Rodrigues2019}. For these reasons, a discrete number of paths must be considered when designing for LSAM, as shown in Fig.~\ref{fig:WAAM_Constraints_d}. Thus, designs with resolution and geometrical features compatible with the nominal production are obtained.

Ways to fully ensure the above mentioned compatibility between design and process are not yet available. Only recently, few studies have appeared that address this gap. \citet{Carstensen2020} considers the size of the deposition nozzle in the TO problem using the formulation that embeds discrete particles in a continuous domain \citep{Guest2015}. The author proposes a formulation that binds the discrete particles to produce a continuous material bead. Numerous examples in 2D are reported, which at first sight meet the imposed resolution of the desired AM process. However, the method binds the particles in a orthogonal pattern, which makes it difficult to generate inclined structural members, therefore, the method is not able to fully exploit the form freedom provided by the AM process. The work of \citet{vantyghem2020} takes a pragmatic approach to deal with the low resolution issue imposed by 3D Concrete Printing. The authors post-process 2D topology-optimized solutions to propose a large-scale three-dimensional design. In the post-processing step, the authors enforce the design details to match the printing resolution of the LSAM process. However, in this manner the optimality of the final design is likely lost.

This article adopts a strategy different from that proposed by \citet{Carstensen2020}, although with the same intent. Consequently, it is aimed to generate topology-optimized designs complying with the low resolution of Large-Scale AM processes, which is typically defined by the size of the deposition nozzle. For this purpose, we propose two methods based on recent developments in minimum and maximum size control in topology optimization \citep{Fernandez2020}. The first method forces the minimum and maximum size of all solid members to be equal to the size of the deposition nozzle. This approach proves to be simple yet effective in producing designs suitable for the desired resolution. However, the maximum size restriction prevents solid members greater than the prescribed nozzle size, which can significantly reduce the achievable design performance. The second method restricts the maximum size of the solid members sufficiently to render a structural skeleton, which can also be interpreted as the deposition path. Through filtering and projection techniques \citep{Wang2011}, the skeleton is dilated according to the nozzle size and thus the actual design is defined. Therefore, a design adapted to the desired resolution is obtained and a low performance penalization is introduced. The proposed strategies are explained, assessed and compared using 2D and 3D numerical examples. Results exhibit designs that conform to the size of the deposition nozzle, allowing optimized topologies to be filled by a discrete number of material beads. Thus, the design requirement shown in Fig.~\ref{fig:WAAM_Constraints_d} is met. Furthermore, for a specific set of printing parameters, deposition paths free of crossovers, T-type connections and sharp corners can be obtained. 

Although this paper addresses a purely geometric limitation that has to do with the size of the deposition nozzle, the obtained results suggest that the nozzle size restrictions proposed for topology optimization improve the printability of optimized components. Other AM constraints that may apply such as critical overhang angles or thermal-induced restrictions are outside the scope of this work. Potential combinations and extensions with other AM constraints are discussed in the Perspectives section towards the end of this manuscript. 

The remainder of this article is organized as follows. Section \ref{sec:2} introduces the formulation of the TO problem that imposes the minimum and maximum size of solid members. Section \ref{sec:3} presents and compares the two methods proposed in this article using a compliance minimization formulation. In addition to 2D examples in the preceding section, Section \ref{sec:5} compares the two methods on a 3D test case. Section \ref{sec:5.5} validates the methods on benchmarks including stress constraints and displacement functions. Finally, Section \ref{sec:6} provides the conclusions and future perspectives of this work.

\section{Base topology optimization formulation} \label{sec:2}

The topology optimization problem is formulated using the density method \citep{Bendsoe1988}. Therefore, the design domains are discretized into $N$ finite elements and their densities are described by $N$ design variables. The value of these design variables range from 0 to 1, where the value 0 represents the void phase and the value 1 represents the solid phase \citep{Bendsoe1989}. To avoid the presence of intermediate densities, we use the Solid Isotropic Material with Penalization (SIMP) interpolation law \citep{Bendsoe1989}. Note that, for simplicity and because the focus of this work is on the nozzle size restriction, anisotropic material behavior is not considered.

It is well-known that SIMP exhibits numerical difficulties, such as the mesh-dependency and the presence of checkerboard patterns \citep{Sigmund1998}. In addition, the formulation does not allow to impose precisely a minimum size control in the solid and void phases \citep{Wang2011}, which is an essential requirement for the methods proposed in this work. For this reason, we adopt the robust formulation proposed by \citet{Sigmund2009}. This formulation considers manufacturing errors that may result in a thinner (eroded) or thicker (dilated) design with respect to the reference one intended for manufacturing (intermediate). These designs are constructed using filtering and projections techniques, which are described below.

The filtering operation utilizes a weighted average of the design variables within a circular (or spherical) domain of radius $r_\mathrm{fil}$ \citep{Bruns2001,Bourdin2001}. The numerical procedure is defined as follows:
\begin{equation} \label{EQ:density_filter}
\tilde{\rho}_i = \frac{\displaystyle\sum_{j=1}^{N}\rho_j \mathrm{v}_j w(\mathbf{x}_i,\mathbf{x}_j)}{\displaystyle\sum_{j=1}^{N} \mathrm{v}_j w(\mathbf{x}_i,\mathbf{x}_j) } \; , 
\end{equation}
where ${\rho}_i$ and $\tilde{\rho}_i$ are the design variable and the filtered variable associated with the finite element $i$, respectively. The volume of element $i$ is denoted as $\mathrm{v}_i$ and the weight of the design variable $\rho_j$ in the definition of $\tilde{\rho}_i$ is denoted by $w(\mathbf{x}_i,\mathbf{x}_j)$. The weighting function depends on the distance between the centroids of elements $i$ and $j$, which are denoted as $\mathbf{x}_i$ and $\mathbf{x}_j$, respectively. Here, the weighting function is defined by the following decreasing linear function:
\begin{equation}
w ( \mathrm{\mathbf{x}}_i,\mathrm{\mathbf{x}}_j) = \mathrm{max}  \left(0 \; , \; 1-\frac{\| \mathrm{\mathbf{x}}_i - \mathrm{\mathbf{x}}_j \|}{r_\mathrm{fil}} \right) \; .
\end{equation}

\begin{figure}
    \hspace*{-40mm}\subfigure[]{\label{FIG:Three_Field_a}}
    \vspace{20mm}\\
	\hspace*{-40mm}\subfigure[]{\label{FIG:Three_Field_b}}
	\vspace{-40mm}\\
	\centering\includegraphics[width=0.6\linewidth]{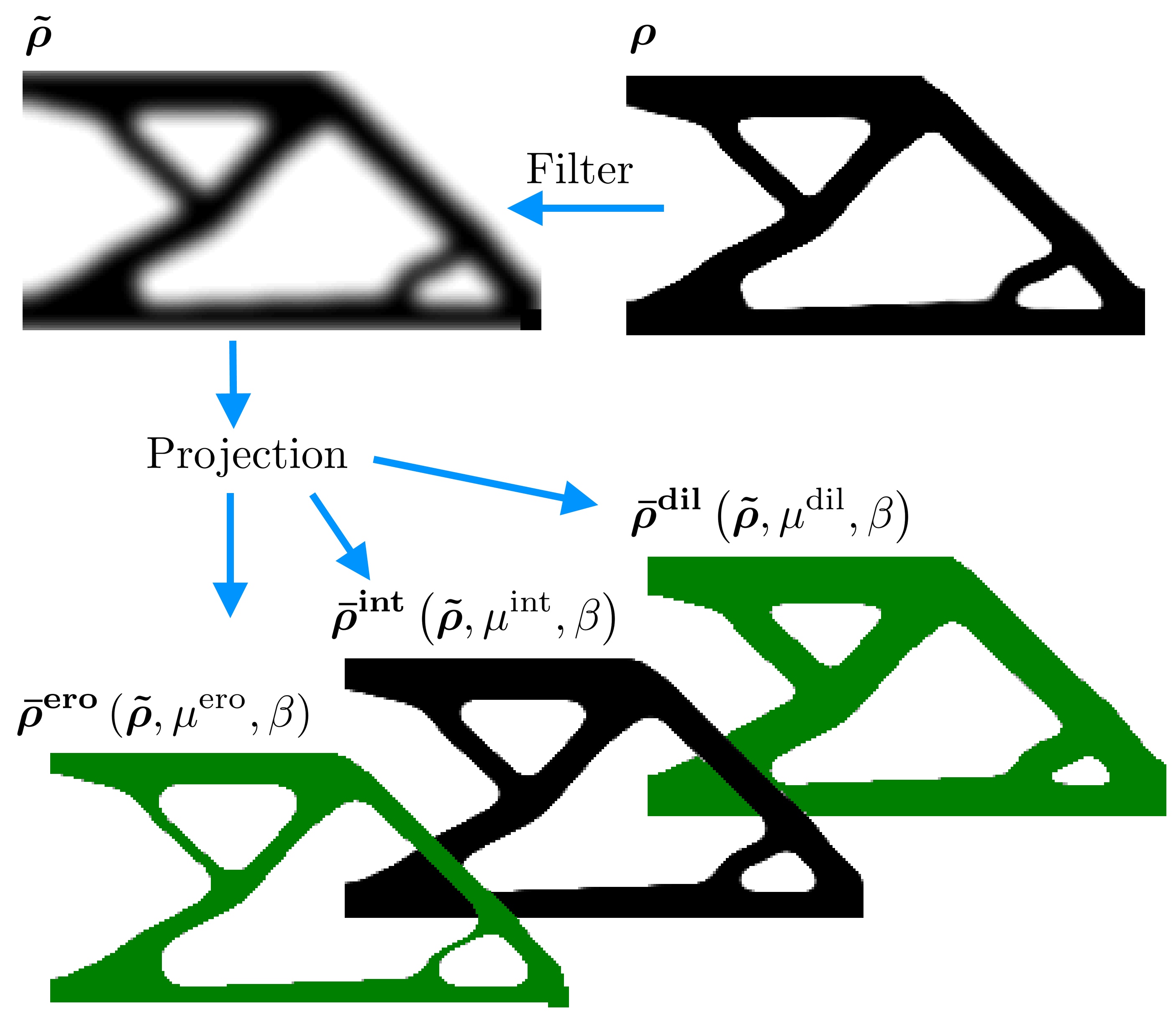}
	\\	
	\caption{A 2D cantilever beam solution (see boundary conditions in Fig.~\ref{FIG:Des_1}) illustrating the three-field scheme. (a) Filtering,  and (b) projection operations to obtain the eroded, intermediate and dilated designs.}	
	\label{FIG:Three_Field}			
\end{figure}

The projection operation transforms the value of all the filtered variables that are greater than a given threshold $\mu$ close to 1, and close to 0 otherwise \citep{Guest2004}. This is done using the following smoothed Heaviside function:
\begin{equation} \label{eq:Heaviside}
\bar{\rho}_i = 
\frac
	{\mathrm{tanh}(\beta\mu)+\mathrm{tanh}(\beta\:(\tilde{\rho}_i-\mu))}
	{\mathrm{tanh}(\beta\mu)+\mathrm{tanh}(\beta\:(        1     -\mu))} \; ,
\end{equation}
\noindent where $\bar{\rho}_i$ is the projected density and $\beta$ is the parameter that controls the steepness of the smoothed Heaviside function \citep{Wang2011}. The parameter $\mu$ controls the projection threshold and can be used to impose a uniform manufacturing error. As can be seen in Fig.~\ref{FIG:Three_Field_b}, the eroded, intermediate and dilated designs are denoted by $\bm{\bar{\rho}}^\mathrm{ero}$, $\bm{\bar{\rho}}^\mathrm{int}$ and $\bm{\bar{\rho}}^\mathrm{dil}$, and they are created selecting the values of thresholds $\mu^\mathrm{ero}$, $\mu^\mathrm{int}$ and $\mu^\mathrm{dil}$, respectively. Fig.~\ref{FIG:Three_Field_b} depicts the eroded and dilated designs colored green to emphasize the fact that the intermediate design is the actual design intended for manufacturing.

\subsection{Minimum size formulation}

\begin{figure}
	\centering
    \includegraphics[width=0.7\linewidth]{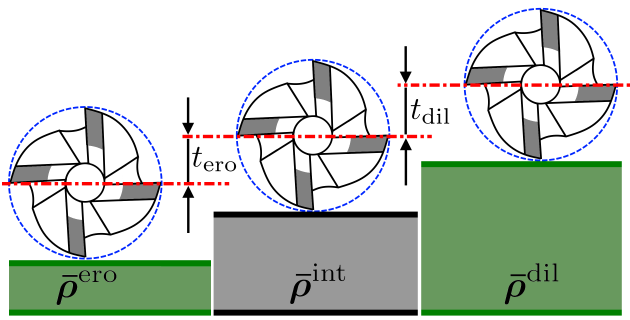}
	\caption{Misplacement of a milling tool. Illustration of the manufacturing error considered in the robust formulation.}	
	\label{FIG:milling}			
\end{figure}

As mentioned above, the eroded and dilated fields can be seen as the outcome product of a manufacturing error. As illustrated in Fig.~\ref{FIG:milling}, the manufacturing error can be caused by the misplacement of a milling tool with respect to the reference line, where $t_\mathrm{ero}$ and $t_\mathrm{dil}$ represent the placement errors.

\citet{Sigmund2009} proposed a robust formulation with respect to manufacturing errors, i.e.~the component must provide high performance even if the optimized design is eventually eroded or dilated during the manufacturing process. For example, considering the compliance minimization problem subject to a volume restriction, the robust formulation reads as follows: 
\begin{align} \label{eq:OPTI_Robust_Design}
	\begin{split}
  		{\min_{\bm{\rho}}} &\quad \max \left(  c(\bm{\bar{\rho}}^{\mathrm{ero}}) \:,\:  c(\bm{\bar{\rho}}^{\mathrm{int}}) \:,\:  c(\bm{\bar{\rho}}^{\mathrm{dil}})\right)				\\
	  	\mathrm{s.t. :} &\quad \mathbf{v}^{\intercal} \bm{\bar{\rho}}^{\mathrm{ero}} \leq V^{\mathrm{ero}} 	\\
	  			&\quad \mathbf{v}^{\intercal} \bm{\bar{\rho}}^{\mathrm{int}} \leq V^{\mathrm{int}}    \\
	  			&\quad \mathbf{v}^{\intercal} \bm{\bar{\rho}}^{\mathrm{dil}} \leq V^{\mathrm{dil}}    \\
	  			&\quad 0 \leq {\rho_i} \leq1 \; ,
	\end{split}
\end{align}
\noindent where $c$ is the compliance, $\mathbf{v}$ is the array containing the elemental volumes, and $V^{\mathrm{ero}}$, $V^{\mathrm{int}}$ and $V^{\mathrm{dil}}$ are the maximum volume of material allowed in the eroded, intermediate and dilated designs, respectively. 

The compliance of each design field is computed as $c=\bm{f}^\intercal \bm{u}$, where $\bm{f}$ is an array containing the external forces and $\bm{u}$ is array containing the nodal displacement of the design field. A linear-elastic response is assumed in the finite element model that yields $\bm{u}$, and the finite element analysis is performed prior to solving the iteration of the optimization problem (approach known as Nested Analysis and Design \citep{Arora2005}).

An interesting outcome of the robust formulation is that the design intended for manufacturing features a minimum size in both the solid and void phases \citep{Wang2011}. The minimum size of the solid is defined by the radius $r_\mathrm{min.Solid}$ of the largest circle (or sphere), that can be inscribed in the thinnest solid member. Conversely, the minimum size of the void is defined by the radius $r_\mathrm{min.Void}$ of the largest circle (or sphere), that can be inscribed in the smallest cavity of the topology. Thus, $r_\mathrm{min.Solid}$ may represent the radius of a deposition nozzle in an AM process, while $r_\mathrm{min.Void}$ may represent the radius of a cylindrical (or spherical) milling tool. 

It is well known that in the robust formulation, the desired minimum length scale is implicitly imposed through the parameters that define the filtering and projection processes \citep{Wang2011}. Thus, $r_\mathrm{min.Solid}$ and $r_\mathrm{min.Void}$ must be defined through $r_\mathrm{fil}$, $\mu^\mathrm{ero}$, $\mu^\mathrm{int}$, and $\mu^\mathrm{dil}$. The relation between the desired minimum length scales and the parameters that impose it can be obtained numerically \citep{Wang2011} or analytically \citep{Qian2013}. To facilitate the description of the proposed methodology, the minimum length scale is fixed at $r_\mathrm{min.Solid}=r_\mathrm{min.Void}$, which is imposed through:
\begin{align} \label{eq:parameters_three_field}
	\begin{split}
  		\mu^\mathrm{ero} &= 0.75 \;\;,\\
  		\mu^\mathrm{int} &= 0.50 \;\;,\\
  		\mu^\mathrm{dil} &= 0.25 \;\;,\\
  		r_\mathrm{fil}   &= 2.0 \: r_\mathrm{min.Solid} \;. \\ 
	\end{split}
\end{align}  
It is important to note that other combinations of length scales and threshold parameters do not change the findings of this work, as long as $0 < \mu^\mathrm{dil} < \mu^\mathrm{int} < \mu^\mathrm{ero} < 1$. 

The eroded design is the most compliant. Therefore, in the compliance minimization problem of Eq.~\eqref{eq:OPTI_Robust_Design}, the intermediate and dilated designs can be excluded from the objective function \citep{Amir2018}. In addition, the volume restriction can be applied to one field only and implicitly restrict the volume of the other two \citep{Sigmund2009}. Thus, the robust compliance minimization problem reduces to the following formulation:
\begin{align} \label{eq:Reference_Opti}
	\begin{split}
  		{\min_{\bm\rho}} &\quad c(\bm{\bar{\rho}}^{\mathrm{ero}}) 			\\
	  	\mathrm{s.t. :} &\quad \mathbf{v}^{\intercal} \bm{\bar{\rho}}^{\mathrm{dil}} \leq V^{\mathrm{dil}}(V^{\mathrm{int}}) 	\\
	  			&\quad 0 \leq {\rho_i} \leq1 \; .
	\end{split}
\end{align}

In the simplified robust topology optimization problem of Eq.~\eqref{eq:Reference_Opti}, the volume constraint is applied to the dilated design. As the intermediate design is the one intended for manufacturing, the desired volume constraint $V^{\mathrm{int}}$ is used to scale the upper bound $V^{\mathrm{dil}}$ that defines the volume constraint \citep{Amir2018}. There are two main reasons for this implicit approach to impose the volume constraint. The first one is the most known in the literature \citep{Amir2018,Fernandez2020,Sigmund2009,Da2019,Wang2011} and it is due to the fact that applying the restriction in the dilated design reduces numerical instabilities and promotes convergence to better optimum. The second reason, which is rarely mentioned in the literature \citep{Trillet2020}, is that the implicit volume restriction allows to involve all 3 design fields in the optimization problem, which is an essential requirement to ensure the minimum size control on the solid and void phases \citep{Trillet2020}.

The minimum compliance optimization problem in Eq.~\eqref{eq:Reference_Opti} is denoted as the \textit{Reference} problem hereafter.

\subsection{Minimum and Maximum size formulation}

\begin{figure}
	\centering
    \includegraphics[width=0.40\linewidth]{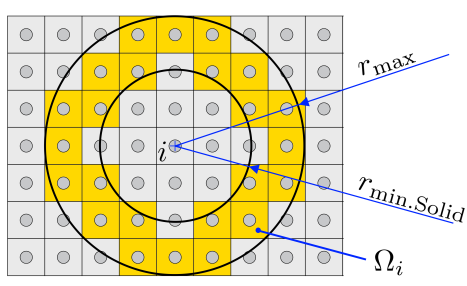}
	\caption{The annular region $\Omega_i$ where the local volume constraint $g_i$ is applied.}	
	\label{FIG:MaxSize_Region}			
\end{figure}

The topology optimization methods accounting for the size of the deposition nozzle require in addition control over the maximum size of the solid members. To this end, the approach proposed by \citet{Fernandez2020} is used. In \citep{Fernandez2020}, the maximum size of a solid feature is imposed using the following local volume restriction \citep{Guest2009}:
\begin{equation} \label{eq:MS_Local_Constraint}
	g_{i}(\bm{\bar{\rho}},\varepsilon,q,\Omega_i) =  \varepsilon - \frac{\displaystyle\sum_{j \in \Omega_i} v_j\:(1-\bar{\rho}_j)^q} {\displaystyle\sum_{j \in \Omega_i} v_j} \leq 0 \quad ,
\end{equation}
\noindent where $g_{i}$ is the volume restriction applied within a region $\Omega_i$. The expression $(1-\bar{\rho}_j)^q$ is a measure of the amount of void within the element $j$, where $q$ is a parameter penalizing intermediate densities to avoid their appearance during the optimization process, similar to the SIMP penalty parameter. Therefore, the volume restriction in Eq.~\eqref{eq:MS_Local_Constraint} computes the fraction of void within the local region $\Omega_i$ and enforces it to be equal or greater than a given fraction $\varepsilon$. The local region $\Omega_i$ is defined as an annulus (or spherical shell), with an outer radius $r_\mathrm{max}$ and inner radius $r_\mathrm{min.Solid}$, as shown in Fig.~\ref{FIG:MaxSize_Region}. Similar to \citep{Fernandez2020,Guest2009}, the void fraction within the local region is set as $\varepsilon=0.05$ and the penalty parameter is set as $q=2$. 

\citet{Fernandez2020} showed that to obtain an intermediate design $\bm{\rho}^\mathrm{int}$ satisfying the desired length scale, i.e.~the minimum member size ($r_\mathrm{min.Solid}^\mathrm{int}$), the minimum cavity size ($r_\mathrm{min.Void}^\mathrm{int}$) and the maximum member size ($r_\mathrm{max}^\mathrm{int}$), it is sufficient to apply the local volume restriction $g_i$ at least in the dilated design. Thus, the compliance minimization problem with minimum and maximum length scale control is defined as:
\begin{align} \label{eq:MaxSize_Opti}
	\begin{split}
  		{\min_{\bm\rho}} &\quad c(\bm{\bar{\rho}}^{\mathrm{ero}}) 			\\
	  	\mathrm{s.t. :} &\quad \mathbf{v}^{\intercal} \bm{\bar{\rho}}^{\mathrm{dil}} \leq V^{\mathrm{dil}}(V^{\mathrm{int}}) 	\\
	  			& \quad \mathrm{G_{ms}}( \bm{\bar{\rho}}^{\mathrm{dil}}) \leq 0 	\\
	  			&\quad 0 \leq {\rho_i} \leq1 \; ,
	\end{split}
\end{align}
The optimization problem in Eq.~\eqref{eq:MaxSize_Opti} is named the \textit{Size-}\linebreak \textit{Constrained} problem hereafter. The $\mathrm{G_{ms}}$ constraint represents the global maximum size restriction that aggregates the local ones ($g_i$). The aggregation is performed with a \textit{p-mean} function, as follows:
\begin{align} \label{eq:GMS}
	\mathrm{G_{ms}} = \left( \frac{1}{N} \sum_{i=1}^N (g_i + 1 - \varepsilon)^p \right)^{1/p} -1 + \varepsilon \leq 0 \; .
\end{align}
\noindent where $p$ is the aggregation exponent and $\varepsilon$ a shift parameter to control aggregation accuracy. Similar to \citep{Fernandez2019,Fernandez2020}, $p$ is set to 100 and $\varepsilon$ to 0.05 in all the examples reported in this work.

\begin{figure}
	\centering
    \includegraphics[width=0.7\linewidth]{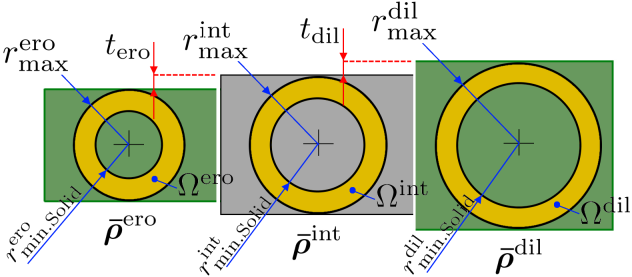}
	\caption{The local $\Omega_i$ in the eroded, intermediate and dilated designs.}	
	\label{FIG:MaxSize_Radius}			
\end{figure}

To apply the local volume constraint $g_i$ in the dilated design, the local region $\Omega_i^\mathrm{dil}$ must be scaled according to the dilation distance $t_\mathrm{dil}$, as illustrated in Fig.~\ref{FIG:MaxSize_Radius}. Thus, the annular region $\Omega_i^\mathrm{dil}$ is defined by the following outer and inner radii:
\begin{align}
	\begin{split}
  		r_\mathrm{max}^\mathrm{dil} & = r_\mathrm{max}^\mathrm{int} + t_\mathrm{dil} \;\;,\\
  		r_\mathrm{min.Solid}^\mathrm{dil} & = r_\mathrm{min.Solid}^\mathrm{int} + t_\mathrm{dil} \;. \\ 
	\end{split}
\end{align}  

Given the filter and projection parameters defining the minimum length scale in Eq.~\eqref{eq:parameters_three_field}, the dilation distance is \citep{Fernandez2020,Trillet2020}:
\begin{equation}
 t_\mathrm{dil} = 0.60 \: r_\mathrm{min.Solid}^\mathrm{int} \;.
\end{equation}

Since the methods presented above are available in the literature, sensitivity analysis is omitted for brevity. Interested readers are referred to \citep{Bendsoe2013} for the sensitivity analysis of the objective function and of the volume restriction, to \citep{Wang2011} for the sensitivity analysis of the eroded, intermediate and dilated designs, and to \citep{Fernandez2020} for the sensitivity analysis of the maximum size restriction.

\subsection{Implementation details}

Given the highly non-linearity of the Heaviside projection and of the aggregation function that builds the maximum size constraint, it is useful to implement a continuation scheme. As done in \citep{Fernandez2020}, the Heaviside parameter $\beta$ is initialized at 1.5 and is increased by 1.5 times up to a maximum of 38. The SIMP penalty parameter is initialized at 1.0 and is increased up to 3.0 in increments of 0.25. The parameter increments are performed every 50 iterations. The stopping criterion of the complete TO process is met either when 450 iterations are reached, or when the maximum change in the design variables between two consecutive iterations is smaller than 0.001. 

The reference and the maximum-size-constrained problems, i.e.~Eqs.~\eqref{eq:Reference_Opti} and \eqref{eq:MaxSize_Opti}, are implemented in free access codes. Here we use the 88-line code \citep{Andreassen2011} and the TopOpt code \citep{Aage2015}. The first one is written in MATLAB aimed at solving 2D problems whereas the latter is a C++ code intended to solve large scale 3D problems. These codes use the density method, the SIMP interpolation scheme and the density filter. Therefore, the Heaviside projection that builds the eroded, intermediate and dilated designs has been added, along with the maximum size restriction and the robust topology optimization formulation. The optimization problems are solved using the Method of Moving Asymptotes (MMA) \citep{Svanberg1987}. 

The reader interested on implementation details is referred to \citep{Fernandez2020}, since the TopOpt code with the corresponding modifications to solve the reference and the maximum size constrained problems are provided. 

\begin{figure}
	\centering
    \includegraphics[width=0.40\linewidth]{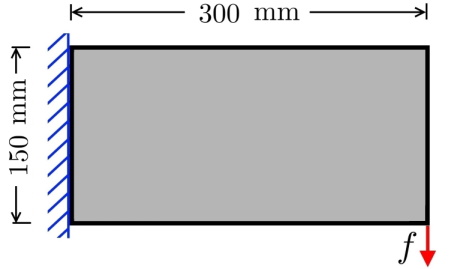}
	\caption{2D cantilever beam design domain.}	
	\label{FIG:Des_1}			
\end{figure}

\section{TO for LSAM} \label{sec:3}

Having introduced the framework to control minimum and maximum size within a topology optimization framework, we proceed with the discussion of the two methods to ensure designs compatible with the intended LSAM process. First, a test problem is introduced by which the methods are explained. Also solutions from the reference optimization problem are discussed to clarify the need for specific design restrictions, after which in subsection \ref{sec:3.1} a method based on a nozzle size constraint is presented. Subsection \ref{sec:3.2} proceeds with the second method, which is based on dilating a maximum size-constrained design.

The proposed strategies are explained using a 2D cantilever beam for compliance minimization. For the sake of clarity, arbitrary yet realistic dimensions are chosen in this test case. The design domain is defined by a rectangle of 300 mm $\times$ 150 mm, as shown in Fig.~\ref{FIG:Des_1}. This domain is discretized into 300 $\times$ 150 quadrilateral finite elements, resulting in square elements of 1 mm length. The optimized design is intended for an AM process equipped with a deposition nozzle of 5 mm diameter. Henceforth, the size of the deposition nozzle is defined by its radius, which is denoted as $r_\mathrm{nozzle}$.  

To provide an insight into the printability of the reported designs, the deposition paths obtained with the PrusaSlicer (v.2.2.0) software are provided. PrusaSlicer \citep{Prusa2020} is a freely available software that generates G-codes from STL files. The software is designed for desktop 3D printers, where the nozzle diameter is approximately 0.4 mm, which is rather small compared to a typical component intended for LSAM machines. Although the nozzle size can be modified to a bigger size, the slicing algorithms of PrusaSlicer (v.2.2.0) do not take into account the low-resolution issue caused by large deposition nozzle sizes. Nonetheless, this software is employed herein as a tool to provide the reader with an idea of the printability of the component when large nozzle sizes are used. In this work, the printability of the optimized designs appears next to the symbol \scalerel*{\includegraphics{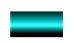}}{B} on the corresponding figures and it represents the percentage of the design that can be filled using a nozzle of a given size. Thus, the printability index (\scalerel*{\includegraphics{Figures/Printability.png}}{B}) is be defined as:
\begin{equation} \label{eq:Printability}
 \mathrm{Printability} :=  \frac{\displaystyle \sum_{i=1}^{n_\mathrm{Layers}} A^\mathrm{filled}_i}{\displaystyle \sum_{i=1}^{n_\mathrm{Layers}}A_i} \times 100 \;\; [\%]  \quad ,
\end{equation}
\noindent where $n_\mathrm{Layers}$ is the number of layers resulting from the slicing procedure, which takes the value 1 in 2D designs. The surface area of the layer desired to be filled by the deposition process is denoted by $A_i$, and the actual surface area filled is denoted by $A^\mathrm{filled}_i$, where $i$ is the index of the $i^\mathrm{th}$ layer.

PrusaSlicer offers different infill patterns, from which the \textit{perimeter} one is chosen. Thus, the slicing algorithm tracks the design perimeter sequentially, which is a common practice in LSAM processes, especially those equipped with large deposition nozzles. However, the software requires at least two deposition paths per solid member, one for each side of the member. Therefore, to ensure that each solid member found in the optimized design contains at least two deposited beads, the minimum size of the solid phase must be twice the size of the nozzle, i.e.~$r_\mathrm{min.Solid}^\mathrm{int} = 2r_\mathrm{nozzle} = 5$ mm. 

Given that the deposition paths follow the contour of the design, it is convenient to control the minimum size of the void phase in order to impose a radius of curvature at the reentrant corners of the design and avoid the difficulties associated to the sharp corners in the deposition path.  Arbitrarily, the minimum radius of curvature is set as $r_\mathrm{min.Void}^\mathrm{int} = 5$ mm, which is imposed with the filter and projection parameters listed in Eq.~\eqref{eq:parameters_three_field}.

Three reference problems (Eq.~\eqref{eq:Reference_Opti}) are solved for the cantilever beam depicted in Fig.~\ref{FIG:Des_1} with different volume restrictions, i.e.~$V^\mathrm{int} = 30\%$, $V^\mathrm{int} = 40\%$ and $V^\mathrm{int} = 50\%$. It is recall that no maximum size constraint is used. The optimized designs and the \textit{perimeter} deposition paths are shown in the first and second rows of Table \ref{TAB:MaxSize}, respectively. The two circles next to each optimized solution indicate the desired length scale. The blue circle represents the minimum size of the void phase while the black circle represents the minimum size of the solid phase. Both have been set to 5 mm. Regarding the deposition paths, the deposited material is shown in light blue and the regions that cannot be filled are shown in yellow. The printability of the optimized design is reported next to the deposition paths. To complement the analysis and discussion, deposition paths obtained with a \textit{rectilinear} filling pattern, and combination of \textit{perimeter} and \textit{rectilinear} are included in the third and fourth rows of Table \ref{TAB:MaxSize}, respectively.

\begin{table}[t!]
\caption{Optimized reference cantilever beams for compliance minimization with minimum size control, for the problem defined in Fig.~\ref{FIG:Des_1}. The circles next to each solution represent the minimum size of the solid (black circle) and void (blue circle) phases. The second row shows the deposition path provided by the PrusaSlicer software using a \textit{perimeter} pattern, the third row using a \textit{rectilinear} pattern, and the fourth row using a combination of \textit{rectilinear} and \textit{perimeter} pattern. Next to each deposition path, the printability (\scalerel*{\includegraphics{Figures/Printability.png}}{B}) is reported, as defined in Eq.~\eqref{eq:Printability}.}
\centering
\begin{tabular}{m{0.2cm} m{4.15cm} m{4.15cm} m{4.15cm}}
	\toprule	    
    & \hspace{14mm} {$V^\mathrm{int} = 30 \%$} & \hspace{14mm} {$V^\mathrm{int} = 40 \%$} & \hspace{14mm} {$V^\mathrm{int} = 50 \%$}  	
	\\
	\cmidrule(r){1-4}
	{\rotatebox{90}{\small{Design}}}
	&   \includegraphics[width=1.0\linewidth]{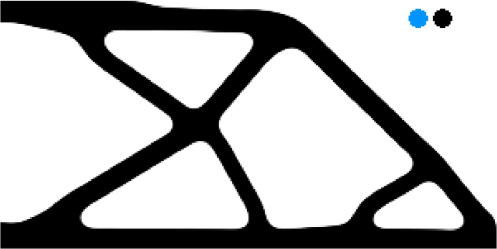}		
	&   \includegraphics[width=1.0\linewidth]{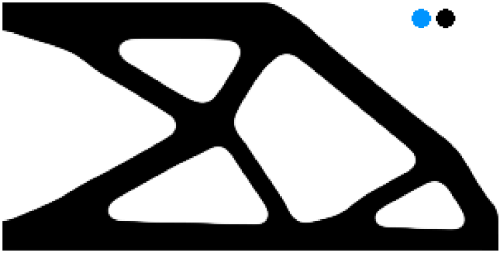}	
	&   \includegraphics[width=1.0\linewidth]{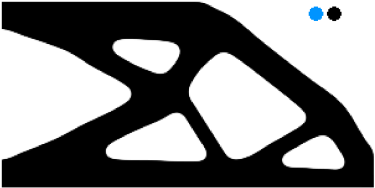}
	\\
	& \hspace{15mm} \footnotesize{c($\bm{\bar{\rho}}^\mathrm{int}$)=124.4}
	& \hspace{15mm} \footnotesize{c($\bm{\bar{\rho}}^\mathrm{int}$)=88.8}
	& \hspace{15mm} \footnotesize{c($\bm{\bar{\rho}}^\mathrm{int}$)=73.2}
	\\
	\cmidrule(r){1-4}
	{\rotatebox{90}{\small{perimeter}}}
	&   \includegraphics[width=1.0\linewidth]{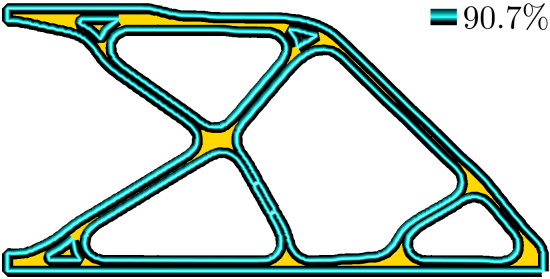}		
	&   \includegraphics[width=1.0\linewidth]{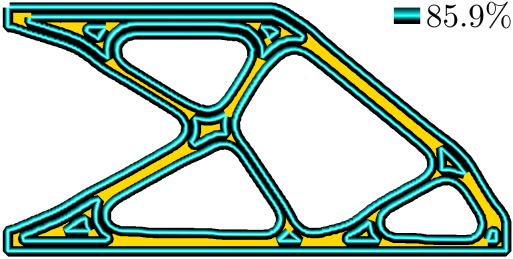}	
	&   \includegraphics[width=1.0\linewidth]{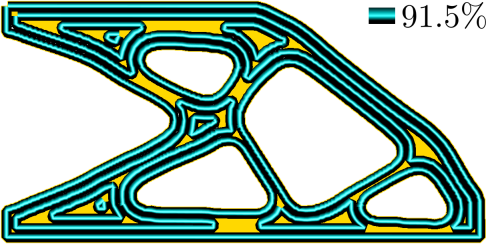}
	\\
	\cmidrule(r){1-4}
	{\rotatebox{90}{\small{rectilinear}}}
	&   \includegraphics[width=1.0\linewidth]{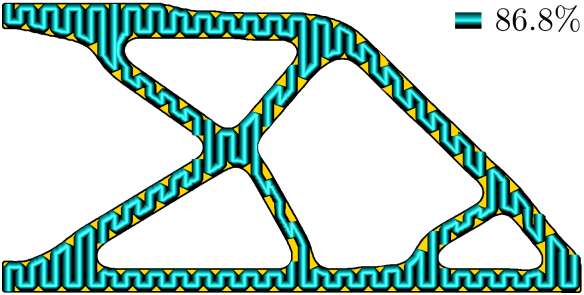}		
	&   \includegraphics[width=1.0\linewidth]{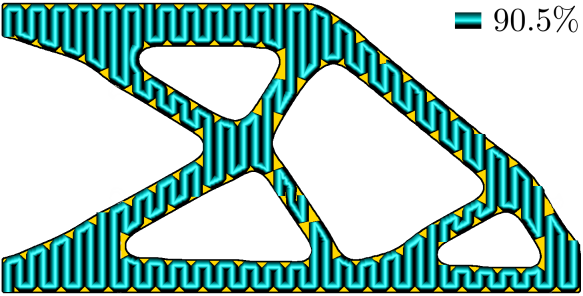}	
	&   \includegraphics[width=1.0\linewidth]{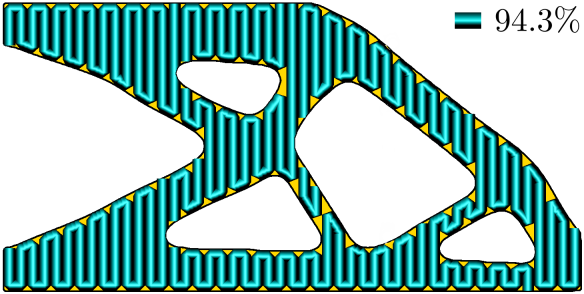}
	\\
	\cmidrule(r){1-4}
	{\rotatebox{90}{\small{combination}}}
	&   \includegraphics[width=1.0\linewidth]{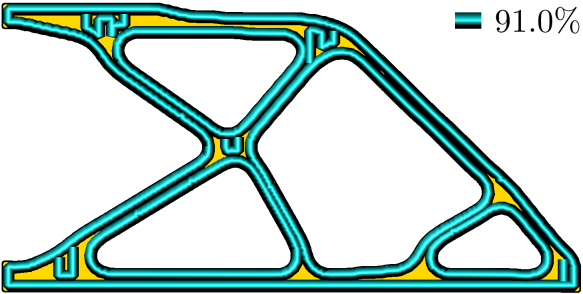}		
	&   \includegraphics[width=1.0\linewidth]{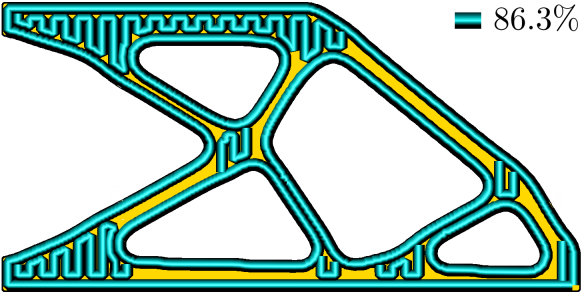}	
	&   \includegraphics[width=1.0\linewidth]{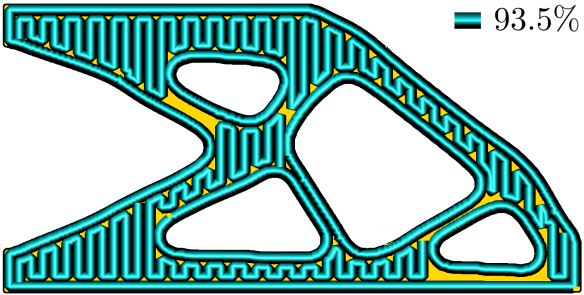}
	\\
	\bottomrule	
\end{tabular}
\label{TAB:MaxSize}	
\end{table}

Considering the \textit{perimeter} infill pattern in Table \ref{TAB:MaxSize}, the reference designs allow fabrication by at least two adjacent beads per solid member, which is attributed to the minimum size control imposed on the topology. However, \textit{perimeter} paths present large unfilled areas that do not match the nozzle size. Better printability can be achieved using other deposition paths or combinations of infill patterns, as shown in the third and fourth rows of Table \ref{TAB:MaxSize}. Even though, it is not possible to guarantee filling all or most of the design during the deposition process, because it is highly likely that numerous design features will not meet the resolution of the deposition nozzle, as this condition is not imposed on the geometry of the reference optimized designs. For example, it can be seen that a \textit{rectilinear} pattern covers more area than the \textit{perimeter} pattern, especially in designs with large volume ($V^\mathrm{int}$ equal to 40$\%$ and 50$\%$) as shown in the third row of Table \ref{TAB:MaxSize}, but it does not cover more than 95$\%$ of the surface due to the small gaps that are present at the outer edge of the design as a result of the numerous path turns. Undoubtedly, the optimal deposition path must consider not only the ability to cover the largest surface area but also other aspects such as the anisotropy of the deposited material, the step-over distance, the number of deposition passes and path continuity, surface finish, and post-machining work among others \citep{Ding2015,Liu2018}. These aspects are not considered in this work and the proposed methods for topology optimization are limited to producing optimized designs tailored to the process resolution in order to facilitate the path planning and guarantee higher printability.

It is also worth noting that better printability can be achieved not only by optimizing the deposition path, but also using a smaller deposition nozzle size in the remaining unfilled areas, by overlapping the beads, or by modifying the feed rate and travel speed. Nevertheless, each of these measures adds a degree of complexity, cost and risk to the process, which is not desired. Instead, the aim of this work is to propose optimized designs suitable with the nozzle size and thus reduce or avoid the need to modify the deposition parameters. For this purpose, the maximum size constraint is used in two different forms, which leads to the two methods proposed in this article.

\subsection{Method 1: Nozzle size constraint} \label{sec:3.1}

The first method conforms the optimized design to the size of the deposition nozzle by tailoring the minimum and maximum size of the solid phase according to the size of the deposition nozzle. Given that in this study it is decided to place two deposited beads per solid member, the minimum and maximum length scales are chosen as follows: 
\begin{equation} \label{eq:MaxSize_Parameters_Nozzle}
	\begin{split}
  		r_\mathrm{min.Solid}^\mathrm{int} &= 2r_\mathrm{nozzle} = 5 \text{mm}  \;\;, \\
  		r_\mathrm{max}^\mathrm{int} & = 2r_\mathrm{nozzle} + \Delta r  = 6 \text{mm}  \;,
	\end{split}
\end{equation}  
\noindent where $\Delta r $ is a small distance to avoid contradictions between the minimum size imposed by the Heaviside projection and the maximum size imposed by the maximum size constraint. In this work, $\Delta r $ is equal to the size of a finite element, which was found to work well. The specific choices in Eq.~\eqref{eq:MaxSize_Parameters_Nozzle} ensure the width of all structural features ranging from 5 mm to 6 mm, i.e.~twice the bead width, which facilitates manufacturability by a LSAM process equipped with a large deposition nozzle fixed in size.

The resulting maximum-size-constrained designs and the corresponding deposition paths are shown in Table \ref{TAB:MaxSize_b}. The desired maximum size of the solid phase is indicated by a black circle next to each solution. Unlike the reference designs, those constrained in maximum feature size possess structural members of uniform width, which leads to fewer unfilled gaps in the designs. The size-constrained members admit two parallel paths, notably without gaps between the deposited beads. In addition, as designs are filled with two parallel beads that follow the surface of the design, no T-joints or path crossovers are obtained. Furthermore, as the minimum size of the void phase is set equal to the radius of the deposition nozzle, no sharp corners are obtained in the deposition path. Note also that the printability of all designs with nozzle size constraint (95.1-98.1$\%$) is superior to that of the reference designs (85.9-91.5$\%$). Therefore, designs constrained in maximum size (Table \ref{TAB:MaxSize_b}) are demonstrated to be much easier to manufacture by LSAM than the reference designs (Table \ref{TAB:MaxSize}). 

\begin{table}
\caption{Optimized cantilever beams and the deposition paths obtained with PrusaSlicer. Design optimized for compliance minimization with simultaneous minimum and maximum size control, for the problem defined in Fig.~\ref{FIG:Des_1}. The difference in compliance with respect to the corresponding reference design (Table \ref{TAB:MaxSize}) is indicated in parentheses. Next to each design, its printability (\scalerel*{\includegraphics{Figures/Printability.png}}{B}) is reported, as defined in Eq.~\eqref{eq:Printability}. The arrows point to areas where the length scale is not fulfilled.}
\centering
\begin{tabular}{m{4.35cm} m{4.35cm} m{4.35cm}}
	\toprule	    
    \hspace{10mm} {$V^\mathrm{int} = 30 \%$} & \hspace{10mm} {$V^\mathrm{int} = 40 \%$} & \hspace{10mm} {$V^\mathrm{int} = 50 \%$}  	
	\\
	\cmidrule(r){1-3}
	    \includegraphics[width=1.0\linewidth]{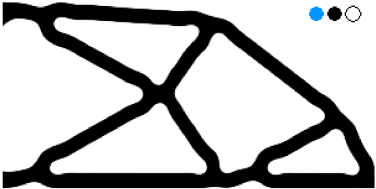}		
	&   \includegraphics[width=1.0\linewidth]{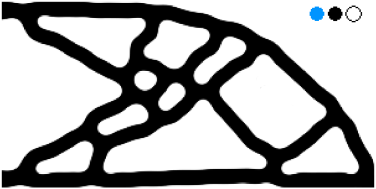}	
	&   \includegraphics[width=1.0\linewidth]{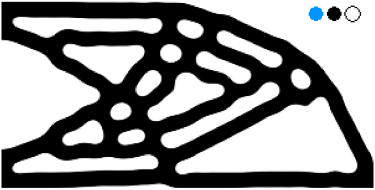}
	\\
	    \includegraphics[width=1.0\linewidth]{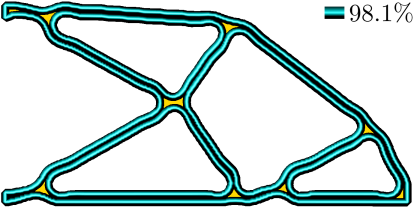}		
	&   \includegraphics[width=1.0\linewidth]{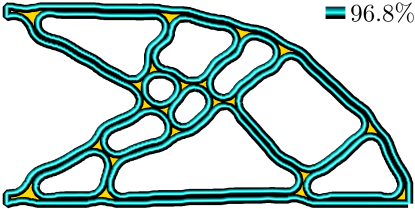}	
	&   \includegraphics[width=1.0\linewidth]{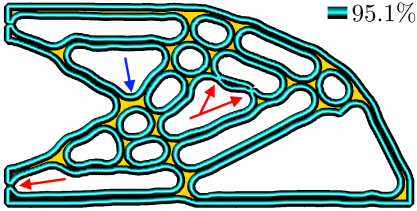}
	\\
	  \hspace{4mm} \footnotesize{c($\bm{\bar{\rho}}^\mathrm{int}$)=136.6 \;\; (+9.8 $\%$)}
	& \hspace{2mm} \footnotesize{c($\bm{\bar{\rho}}^\mathrm{int}$)=119.1 \;\; (+34.1 $\%$)}
	& \hspace{3mm} \footnotesize{c($\bm{\bar{\rho}}^\mathrm{int}$)=94.4  \;\; (+29.0 $\%$)}
	\\
	\bottomrule	
\end{tabular}
\label{TAB:MaxSize_b}	
\end{table}

\begin{figure}	
		\centering	
		\includegraphics[width=0.55\linewidth]{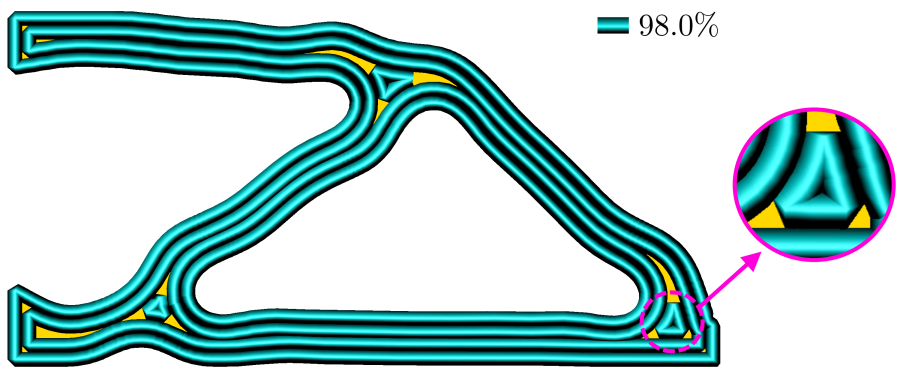}
	\caption{Maximum size-constrained cantilever beam solved for $V^\mathrm{int} = 40\%$ and $r_\mathrm{min.Solid}^\mathrm{int} = 4r_\mathrm{nozzle} = 10 \text{mm}$. The design domain is defined in Fig.~\ref{FIG:Des_1}. The symbol \scalerel*{\includegraphics{Figures/Printability.png}}{B} represents the printability index defined in Eq.~\eqref{eq:Printability}.}			
	\label{fig:MaxSize_x2}
\end{figure}

Despite the benefits introduced by the maximum size restriction, some drawbacks can be noted. For instance, small unfilled zones can be observed at the intersection of two or more solid members. There, the \textit{perimeter} deposition path exposes triangular unfilled regions, which represent about $2\%$ of the design obtained with $V^\mathrm{int} = 30 \%$ in Table \ref{TAB:MaxSize_b}. Due to these unfilled zones, it is expected that performance of the manufactured component will be different from that of the optimized design, although the performance reduction is not quantified in this study. Regarding this geometrical drawback at the intersection of two or more solid members, we point out that the printability of such zones depends on the intended AM process and the deposition path. For example, if the nozzle size is chosen to be equal to the width of the solid members, the intersection between solid members would represent a T-joint or a crossover of paths. In such cases, it may be advantageous to have an intersection larger than the prescribed size since, in practice, it is difficult to achieve a joint with high dimensional accuracy \citep{Venturini2016}. 

The size-constrained design in Table \ref{TAB:MaxSize_b} for a volume restriction of $V^\mathrm{int}=50 \%$ exhibits zones that do not meet the desired minimum size, such as those indicated by the red arrows in the deposition path in Table \ref{TAB:MaxSize_b}. In addition, given that the global maximum size constraint $\mathrm{G_{ms}}$ is defined by a \textit{p-mean} aggregation function, it is possible to find parts of the topology that do not meet the imposed maximum size, as the one highlighted by the blue arrow in Table \ref{TAB:MaxSize_b}. These undesirable features typical of a local optimum can be reduced by adjustment of parameters. For instance, by increasing the $p$ exponent of the aggregation function and increasing the value of the penalty parameters ($\beta$ and SIMP) more slowly, i.e.~by adapting the continuation method to involve a larger number of iterations \citep{Fernandez2020}. 

It is worth noting that the choice of placing two material beads is to make use of the \textit{perimeter} deposition path of PrusaSlicer, which is used for illustrative purposes. That is, the proposed method can be used to place any number of beads along the structural members by choosing the maximum and minimum size of the solid phase according to the desired number of beads. For illustrative purposes, a cantilever beam is optimized taking into account 4 material beads along the structural members. Since an even number of beads is assumed, the \textit{perimeter} pattern can be used to illustrate the deposition path, which is shown in Fig.~\ref{fig:MaxSize_x2}. It is noticeable that a large part of the design is filled with 4 material beads arranged in a parallel fashion, and the remaining gaps at the intersection of solid members are filled with triangular patterns, which may present difficulties when printing due to the sharp corners of the trajectory, as outlined in Fig.~\ref{fig:MaxSize_x2}. Therefore, positioning two material beads per solid member may be a preferable strategy when using maximum size constraints. 

Certainly, one of the most significant drawbacks of the maximum size restriction is the performance reduction. It can be seen that the size-constrained design obtained with $V^\mathrm{int}=40\%$ has a $34\%$ higher compliance compared to the reference design. Requiring all structural members to have the same width of, in this case, two beads, is a rather strong design restriction. Allowing for design features to have varying integer multiples of adjacent beads in various places throughout the structure would also result in a manufacturable design, while offering more design freedom. To conform the optimized design to the size of the deposition nozzle and to avoid a significant reduction on the structural performance, the following subsection presents a second strategy, which is also based on maximum size constraints.

\subsection{Method 2: Skeleton-based deposition paths} \label{sec:3.2}

In the method discussed above, the chosen minimum and the maximum length scales dictate structural members to have uniform width, which facilitates the generation of a feasible deposition path. However, the structural members remain separated from each other, which negatively affects the structural performance. This subsection shows that the separation between members can be eliminated through a dilation projection while retaining the manufacturability. The proposed method is based on the numerical study presented in \citep{Fernandez2020}.    

\citet{Fernandez2020} revealed the behaviour of the maximum size constraint in the context of the robust formulation by restricting the size of  the eroded, intermediate and dilated designs separately. The authors concluded that in order to impose the desired minimum and maximum length scales, it is necessary to restrict at least the dilated design. Part of the numerical study developed in \citep{Fernandez2020} is reproduced in Table \ref{TAB:Strategy} in order to highlight the property in interest for our work. To this end, the maximum-size-constrained problem in Eq.~\eqref{eq:MaxSize_Opti} is solved but with variations on the field where the maximum size constraint is applied. The first and second rows of the table show the result obtained when the maximum size restriction is applied to the intermediate and to the dilated design, respectively. The local region $\Omega_i$ is shown next to the field in which the maximum size constraint is applied, while the desired minimum length scale is shown by the blue and black circles next to the intermediate design. As concluded in \citep{Fernandez2020}, the minimum and maximum length scales are met only when $\mathrm{G_{ms}}( \bm{\bar{\rho}}^{\mathrm{dil}})$ is included in the topology optimization problem. However, in this work we emphasize another important observation.

\begin{table}
\caption{Optimized designs obtained with maximum size constraints applied on different design fields. Circles next to the intermediate design illustrate the desired minimum size in the void phase (blue circle), in the solid phase (black circle), and the maximum size in the solid phase (black ring).}
\centering
\begin{tabular}{m{1.6cm} m{3.7cm} m{3.7cm} m{3.7cm}}
	\toprule	    
     Constraint & \hspace{17mm} \large{$\bm{\bar{\rho}}^\mathrm{ero}$} & \hspace{17mm} \large{$\bm{\bar{\rho}}^\mathrm{int}$} & \hspace{17mm} \large{$\bm{\bar{\rho}}^\mathrm{dil}$}  	
	\\
	\cmidrule(r){1-4}
	{$\mathrm{G_{ms}}( \bm{\bar{\rho}}^{\mathrm{int}})$}
    &   \includegraphics[width=1.0\linewidth]{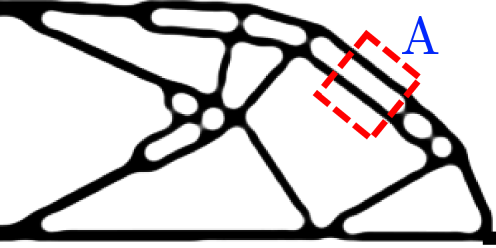}		
	&   \includegraphics[width=1.0\linewidth]{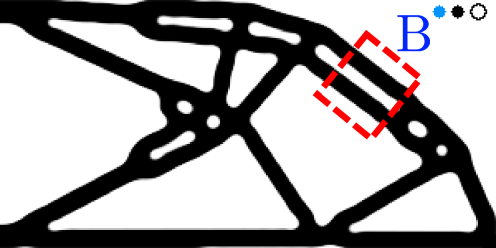}	
	&   \includegraphics[width=1.0\linewidth]{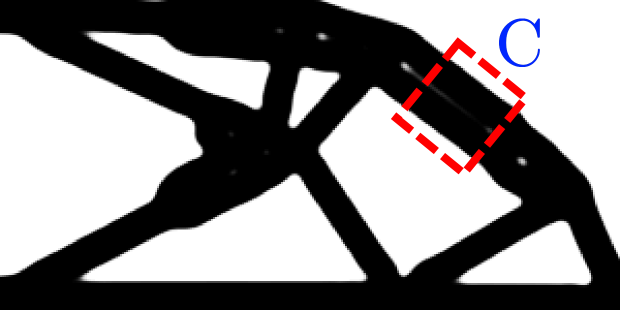}
	\\
	\cmidrule(r){1-4}
	{$\mathrm{G_{ms}}( \bm{\bar{\rho}}^{\mathrm{dil}})$}
    &   \includegraphics[width=1.0\linewidth]{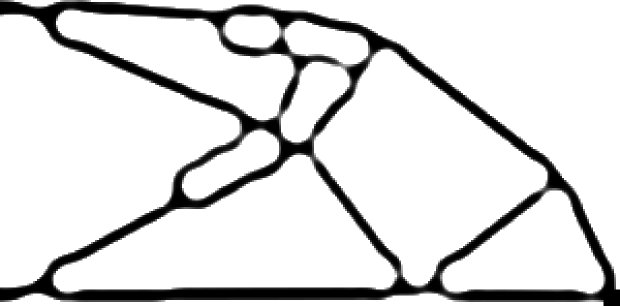}		
	&   \includegraphics[width=1.0\linewidth]{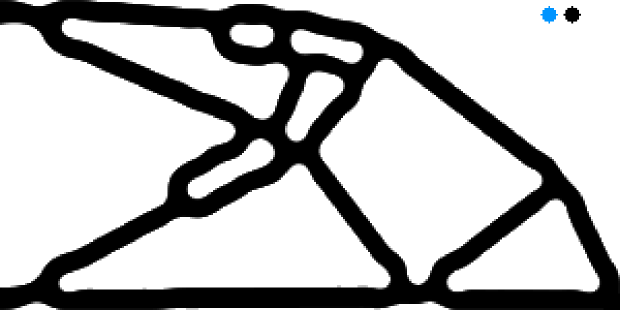}	
	&   \includegraphics[width=1.0\linewidth]{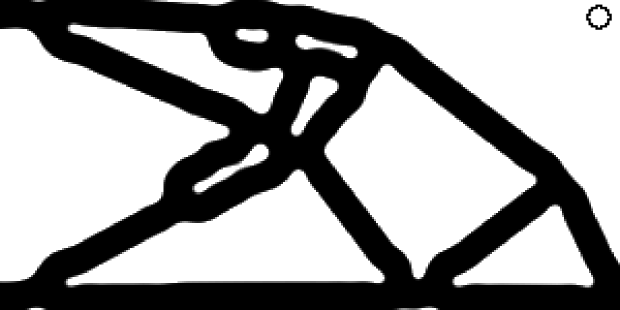}
	\\
	\bottomrule	
\end{tabular}
\label{TAB:Strategy}	
\end{table}

\begin{figure}
	\centering
    \includegraphics[width=0.6\linewidth]{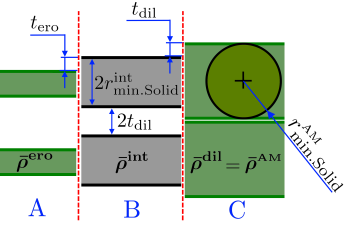}
	\caption{Illustration of the bonding of bars resulting from the dilation projection. Labels A, B and C represent the zones highlighted in Table \ref{TAB:Strategy}. The zone C emphasizes the fact that the design to be manufactured is the dilated one, for which it is now defined as $\bm{\bar{\rho}}^{\text{\tiny{AM}}}$.}	
	\label{FIG:Merging}			
\end{figure}

The first row of Table \ref{TAB:Strategy} suggests that if the maximum size constraint is applied only in the intermediate design, the dilated design does not feature a maximum member width, since the dilation projection closes the cavities present in the intermediate design, as illustrated in Fig.~\ref{FIG:Merging}. Under this formulation, the eroded and intermediate designs can be interpreted as the structural skeleton or the deposition lines of an AM process, and the dilated design would be the component obtained after deposition. For suitably chosen length scales, the bonding of adjacent beads can be represented in this way. This observation allows us to formulate the following topology optimization problem:
\begin{align} \label{eq:Reference_Opti_AM}
	\begin{split}
  		{\min_{\rho}} &\quad  \alpha \: c(\bm{\bar{\rho}}^{\text{\tiny{AM}}}) + (1-\alpha) \: c(\bm{\bar{\rho}}^{\mathrm{ero}}) 			\\
	  	\mathrm{s.t. :} &\quad \mathbf{v}^{\intercal} \bm{\bar{\rho}}^{\text{\tiny{AM}}} \leq V^{\text{\tiny{AM}}} 	\\
	  			& \quad \mathrm{G_{ms}}( \bm{\bar{\rho}}^{\mathrm{int}}) \leq 0 	\\
	  			&\quad 0 \leq {\rho_i} \leq1 \; ,
	\end{split}
\end{align}
\noindent where $\bm{\bar{\rho}}^{\text{\tiny{AM}}}$ represents a dilated projection and the design to be manufactured. Henceforth, this optimization problem \eqref{eq:Reference_Opti_AM} is denoted as the \textit{AM-constrained} problem.

The AM-constrained problem contains the sum of two compliances in the objective function. The compliance of the eroded design serves to impose the minimum size of the solid members \citep{Wang2011}, while the compliance of the dilated design is to ensure a good performance of the component intended for manufacturing. The compliances are weighted by $\alpha$, where $0 \leq \alpha \leq 1$. To define $\alpha$, the following must be considered. If the eroded compliance is prioritized ($\alpha \approx 0$), the minimum size of the solid phase is guaranteed, but as insignificant emphasis is placed on the performance of the printed structure, the bars of the AM design could still be separated by a very thin layer of gray elements (as seen in zone labeled as C in Table \ref{TAB:Strategy}), which could impair structural performance. In addition, if a large volume of material is allowed in the design domain, floating structures can arise in the dilated design \citep{Fernandez2020}. On the other hand, if the dilated compliance is prioritized ($\alpha \approx 1$), the bonding of bars is promoted but the minimum size control is not ensured. If the compliances are weighted equally ($\alpha = 0.5$), a topology that does not meet the desired minimum size and/or that does not completely merge the bars could be obtained. To avoid these issues, the weighting parameter $\alpha$ is defined with a continuation scheme. At the beginning of the optimization problem, the minimum size is prioritized, and as the optimization progresses, the structural performance is prioritized. Thus, $\alpha$ can be defined by the following function:
\begin{equation}
\alpha = \frac{\mathrm{Iter}-1}{\mathrm{Iter_{max}}-1} \; ,
\end{equation} 
\noindent where $\mathrm{Iter_{max}}$ is the maximum number of iterations defined for the continuation scheme and $\mathrm{Iter}$ is the current iteration number. Although this continuation scheme causes the objective to change in every iteration, it was found to have no detrimental effect on convergence, and has been used successfully in all the cases considered in this study.

To select appropriate length scales and obtain designs suitable for Large-Scale AM processes, i.e.~members composed of an integer number of parallel beads, the following must be considered. Given that the design intended for manufacturing is $\bm{\bar{\rho}}^{\text{\tiny{AM}}}$, the desired minimum size in the solid phase is $r_\mathrm{min.Solid}^\text{\tiny{AM}}$, as illustrated in Fig.~\ref{FIG:Merging}. Therefore, the set of projection and filtering parameters ($\mu^\text{ero}$, $\mu^\mathrm{int}$, $\mu^\text{\tiny{AM}}$ and $r_\mathrm{fil}$) must be found for a length scale defined in the dilated design. The information available in the literature to obtain the filtering and projection parameters are meant for a length scale defined in the intermediate design \citep{Wang2011,Fernandez2020}, but it can be used here knowing that the length scale of the intermediate and dilated designs are related by the dilation distance, as follows:  
\begin{equation} \label{eq:rMinAM_tdil}
r_\mathrm{min.Solid}^\text{\tiny{AM}} = r_\mathrm{min.Solid}^\text{\tiny{int}} + t_\mathrm{dil} 
\end{equation} 

\begin{figure}
	\centering
	\subfigure[Graph to obtain $r_\mathrm{min.Solid}^\mathrm{int}(r_\mathrm{fil})$]{
    		\includegraphics[width=0.47\linewidth]{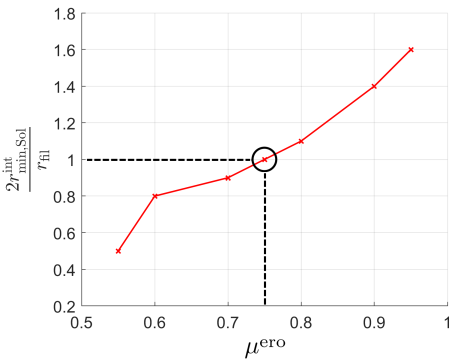}			
		\label{FIG:Projection_Parameters_b}	
	}
	~
	\subfigure[Graph to obtain $t_\mathrm{dil}(r_\mathrm{min.Solid}^\mathrm{int})$]{
    		\includegraphics[width=0.47\linewidth]{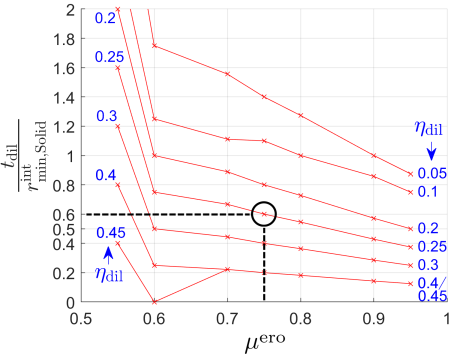}			
		\label{FIG:Projection_Parameters_a}	
	}			
	\\	
	\caption{Graphs relating the projection thresholds, the filter radius, the minimum size of the solid phase in the intermediate design and the dilation distance. The highlighted points are used in Eqs.~\eqref{eq:rMinAM_tdil_3} and \eqref{eq:rMinAM_tdil_4}. The graphs have been constructed with the MATLAB script provided in \citep{Trillet2020}, running the code as \texttt{NumericalSolution(10,38)}.}	
	\label{FIG:Projection_Parameters}			
\end{figure}

Taking into account that the following set of thresholds is adopted:
\begin{align}  \label{eq:rMinAM_tdil_2}
  		[\mu^\text{\tiny{AM}},\; \mu^\mathrm{int},\; \mu^\mathrm{ero}]  & =  [0.25 \;,\; 0.50 \;,\; 0.75]  \;,
\end{align}
\noindent it only remains to find the relationships between $r_\mathrm{min.Solid}^\text{\tiny{int}}$, $t_\mathrm{dil}$ and $r_\mathrm{fil}$, and to replace these relationships in Eq.~\eqref{eq:rMinAM_tdil} to obtain the filter radius leading to $r_\mathrm{min.Solid}^\text{\tiny{AM}}$. To assist the reader in understanding the procedure for obtaining the required relationships, we include in Fig.~\ref{FIG:Projection_Parameters} graphs relating the projection parameters and the minimum sizes of the intermediate design. For the generation of these graphs, the MATLAB code provided in \citep{Trillet2020} has been run as \texttt{NumericalSolution(10,38)}. The input parameter "10" denotes a representative size of the filter radius (given in finite elements) simply needed to reduce rounding errors resulting from the finite element discretization. "38" represents the maximum value of the smoothed Heaviside projection parameter. The procedure to obtain the filter radius for a given nozzle size is detailed hereafter.

The minimum size of the solid phase in the intermediate design ($r_\mathrm{min.Solid}^\text{\tiny{int}}$) can be expressed as a function of the filter radius using Fig.~\ref{FIG:Projection_Parameters_b}, which yields to:  
\begin{align}  \label{eq:rMinAM_tdil_3}
	 r_\mathrm{min.Solid}^\text{\tiny{int}} & = 0.5 r_\mathrm{fil} 
\end{align}

The dilation distance can be expressed as a function of the filter radius using Fig.\ref{FIG:Projection_Parameters_a}, resulting in:
\begin{align} \label{eq:rMinAM_tdil_4}
	 t_\mathrm{dil} & =  0.60 r_\mathrm{min.Solid}^\text{int}
\end{align}

Then, using Eqs.~\eqref{eq:rMinAM_tdil_4} and \eqref{eq:rMinAM_tdil_3} in Eq.~\eqref{eq:rMinAM_tdil}, the filter radius as a function of the desired minimum length scale is obtained:
\begin{align} \label{eq:rMinAM_tdil_5}
	 r_\mathrm{fil} & = 1.25 \: r_\mathrm{min.Solid}^\text{\tiny{AM}}
\end{align}

The procedure explained in Eqs.~\eqref{eq:rMinAM_tdil_3}-\eqref{eq:rMinAM_tdil_5} can be used to obtain the filter radius ($r_\mathrm{fil}$) as a function of the desired length scale ($r_\mathrm{min.Solid}^\text{\tiny{AM}}$) for any other set of thresholds ($\mu^\text{\tiny{AM}}$, $\mu^\mathrm{int}$ and $\mu^\mathrm{ero}$) that may be required for a specific reason.

It is emphasized that the relationships between parameters in Eqs.~\eqref{eq:rMinAM_tdil}-\eqref{eq:rMinAM_tdil_5}, are obtained from the graphs of Fig.~\ref{FIG:Projection_Parameters}, which are built from a numerical method that assumes a 1D design domain \citep{Wang2011} and a filter radius of the order of 10 finite elements. For this reason, it is expected that the relationships provided by Fig.~\ref{FIG:Projection_Parameters} will not be accurately met in practice. A major source of error comes from rounding the number of finite elements, especially when small distances (close to one or two finite elements) are involved \citep{Trillet2020}. This rounding error explains the improved behavior of the method when using a slightly smaller dilation projection threshold, which in our case is 0.20 instead of 0.25. Taking into account this observation, the set of parameters that are used hereafter in the AM-constrained problem are:   
\begin{align} 
	\begin{split}
		r_\mathrm{fil} & = 1.11 \: r_\mathrm{min.Solid}^\text{\tiny{AM}} \\
  		\mu^\mathrm{ero} & = 0.75 \\
  		\mu^\mathrm{int}  & = 0.50 \\
  		\mu^\text{\tiny{AM}} & =  0.20 \;. \\
	\end{split}
\end{align}
\noindent where the factor 1.11 comes from the equations $r_\mathrm{fil}$ = $2 r_\mathrm{min.Solid}^\text{int}$ and $t_\mathrm{dil}$ = $0.80 r_\mathrm{min.Solid}^\text{int}$, which are obtained from Figs.~\ref{FIG:Projection_Parameters_a} and \ref{FIG:Projection_Parameters_b}, respectively. The reason why $\mu^\text{\tiny{AM}} =  0.20$ performs better than $\mu^\text{\tiny{AM}} =  0.25$ is described in the following. It is recalled that in the illustrative example, the minimum size of the solid phase ($r_\mathrm{min.Solid}^\text{\tiny{AM}}$) has been set equal to 5 finite elements. When using a projection $\mu^\text{\tiny{AM}} =  0.25$, the dilation distance is $t_\mathrm{dil}=0.6 \times 0.5 \times 1.25 \times r_\mathrm{min.Solid}^\text{\tiny{AM}} = 1.88$ finite elements. On the other hand, if $\mu^\text{\tiny{AM}} =  0.20$ is used, the dilation distance is $t_\mathrm{dil}=0.8 \times 0.5 \times 1.1 \times r_\mathrm{min.Solid}^\text{\tiny{AM}} = 2.20$ finite elements. Numerically, 1.88 results in one finite element (0.88 elements of rounding error), and 2.2 results in two finite elements (0.2 elements of rounding error). That is, a projection using  $\mu^\text{\tiny{AM}} =  0.20$ induces a smaller rounding error and produces a dilation distance one finite element larger than the one obtained with $\mu^\text{\tiny{AM}} =  0.25$, which results in a better bonding between members in the dilated projection. The reader intending to modify the projection thresholds or imposed length scale is recommended to make use of the numerical method provided in \citep{Trillet2020}, as it entails smaller rounding errors and thus less work in parameter tuning compared to the analytical method provided in the cited work.

Now that the formulation of the topology optimization problem has been presented and the projection and filtering parameters have been defined, the proposed method is demonstrated on the cantilever beam presented in the previous section. As the minimum size of the solid phase is set to twice the size of the nozzle, $r_\mathrm{fil}= 1.11 \: \times 2 \times r_\mathrm{nozzle} = 5.6$ mm. The eroded, intermediate and AM designs, along with the deposition paths are shown in Table \ref{TAB:AM_Results}.

Similar to the maximum-size-constrained designs reported for the first method, the AM-constrained designs do present small unfilled gaps, which are mostly situated at the intersection of solid members. In terms of printability, the designs perform similarly as well. However, the AM-constrained designs exhibit better performance on the face of increased design freedom. The difference is most noticeable when large volumes of material are allowed in the topology, which highlights the need to prefer the AM-constrained optimization problem over the maximum-size-constrained one. 

\begin{table}
\caption{Optimized designs provided by the AM-constrained topology optimization problem. The difference in compliance with respect to the reference design is indicated in parentheses. The percentage next to the symbol \scalerel*{\includegraphics{Figures/Printability.png}}{B} represents the printability index defined in Eq.~\eqref{eq:Printability}.}
\centering
\begin{tabular}{|m{0.5cm}| m{4.0cm}| m{4.0cm}| m{4.0cm}|}
	\toprule	    
    & \hspace{10mm} {$V^\text{\tiny{AM}} = 30 \%$} & \hspace{10mm} {$V^\text{\tiny{AM}} = 40 \%$} & \hspace{10mm} {$V^\text{\tiny{AM}} = 50 \%$}  	
	\\
	\cmidrule(r){1-4}
	{\large{$\bm{\bar{\rho}}^\mathrm{ero}$}}
    &   \includegraphics[width=1.0\linewidth]{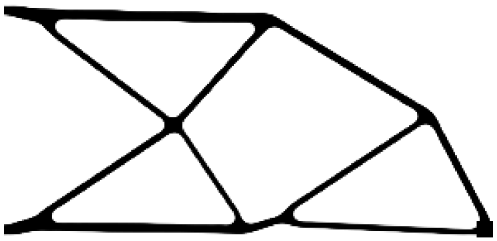}		
	&   \includegraphics[width=1.0\linewidth]{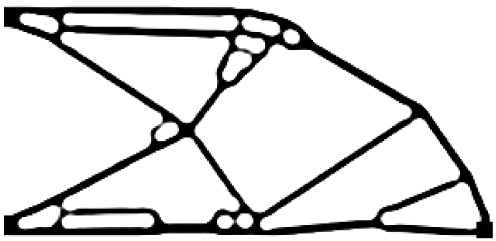}	
	&   \includegraphics[width=1.0\linewidth]{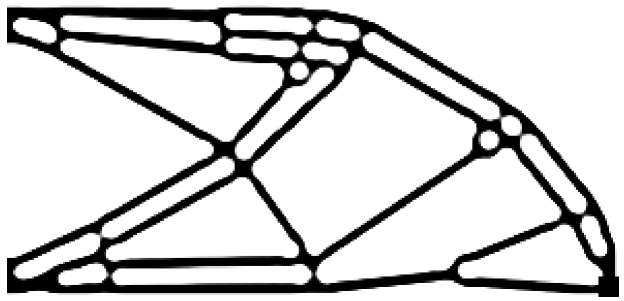}
	\\
	{{$\bm{\bar{\rho}}^\mathrm{int}$}}
	&   \includegraphics[width=1.0\linewidth]{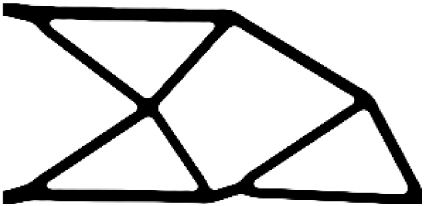}		
	&   \includegraphics[width=1.0\linewidth]{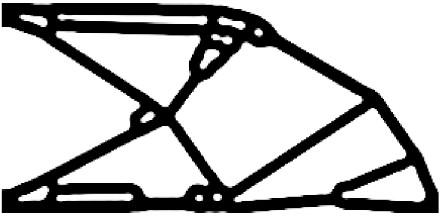}	
	&   \includegraphics[width=1.0\linewidth]{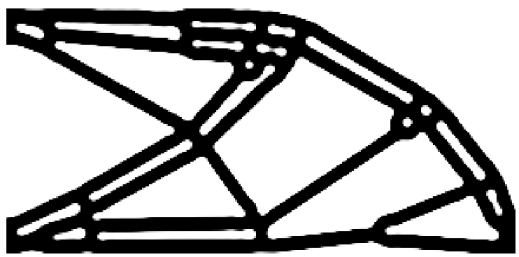}
	\\
	{{$\bm{\bar{\rho}}^\text{\tiny{AM}}$}}
	&   \includegraphics[width=1.0\linewidth]{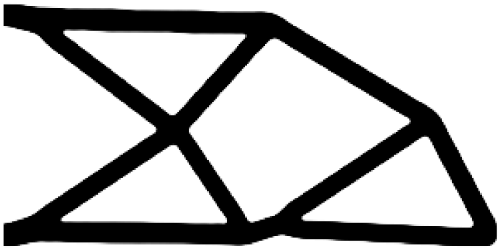}		
	&   \includegraphics[width=1.0\linewidth]{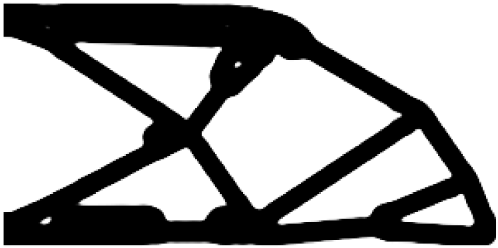}	
	&   \includegraphics[width=1.0\linewidth]{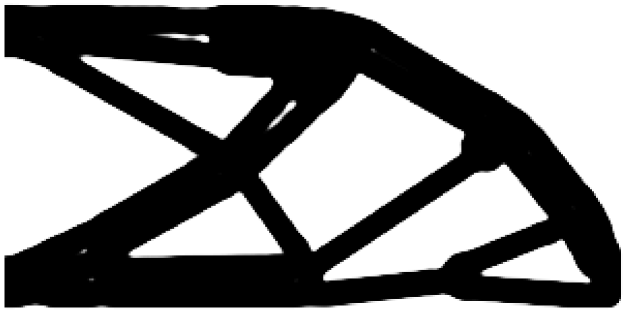}
	\\
	& \hspace{2mm} \footnotesize{c($\bm{\bar{\rho}}^\text{\tiny{AM}}$)=132.3 \;(+6.4$\%$)}
	& \hspace{2mm} \footnotesize{c($\bm{\bar{\rho}}^\text{\tiny{AM}}$)=101.8 \;(+14.6$\%$)}
	& \hspace{2mm} \footnotesize{c($\bm{\bar{\rho}}^\text{\tiny{AM}}$)=80.2  \;(+9.6$\%$)}
	\\ 
	\cmidrule(r){1-4}
	{\rotatebox{90}{{Deposition Path}}}
    &   \includegraphics[width=1.0\linewidth]{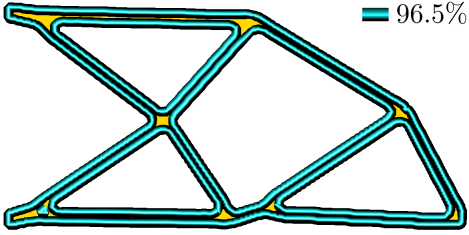}		
	&   \includegraphics[width=1.0\linewidth]{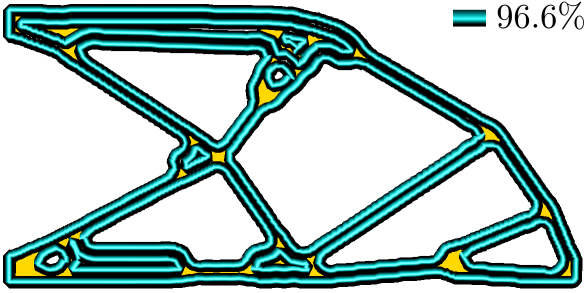}	
	&   \includegraphics[width=1.0\linewidth]{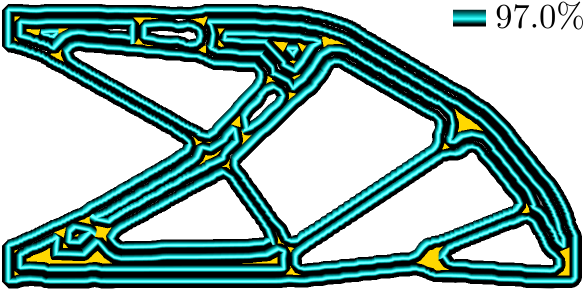}
	\\
	\bottomrule	
\end{tabular}
\label{TAB:AM_Results}	
\end{table}

For a small amount of material, such as $V=30\%$, the reference (Table \ref{TAB:MaxSize}), the size-constrained (Table \ref{TAB:MaxSize_b}) and the AM-constrained (Table \ref{TAB:AM_Results}) optimization problems generate similar topologies. However, for large volumes of material, the topologies can be very different and therein lies the benefit of the AM-constrained formulation. For example, for $V=50\%$, the $\bm{\bar{\rho}}^\text{\tiny{AM}}$ design presents zones where two material beads are placed next to each other to form a bulky zone. As the volume of material is increased in the design space, more parallel material beads merge, as shown in the fourth column of Table \ref{TAB:AM_Results}. 

\begin{table}
\caption{AM-constrained designs for different nozzle sizes and $V^\text{\tiny{AM}}=60\%$. The firs row shows the optimized designs. The second row shows the eroded designs superimposed on the AM-designs. The third row shows the deposition paths obtained with PrusaSlicer (v.2.2.0). The percentage next to the symbol \scalerel*{\includegraphics{Figures/Printability.png}}{B} represents the printability index defined in Eq.~\eqref{eq:Printability}.}
\centering
\begin{tabular}{|m{0.5cm}| m{4.0cm}| m{4.0cm}| m{4.0cm}|}
	\toprule	    
    & \hspace{4mm} {$r_\mathrm{nozzle}=2.5$ mm} & \hspace{4mm} {$r_\mathrm{nozzle}=4.1$ mm} & \hspace{4mm} {$r_\mathrm{nozzle}=5.5$ mm}  	
	\\
	\cmidrule(r){1-4}
	{\rotatebox{90}{{$\bm{\bar{\rho}}^\text{\tiny{AM}}$}}} 
    &   \includegraphics[width=1.0\linewidth]{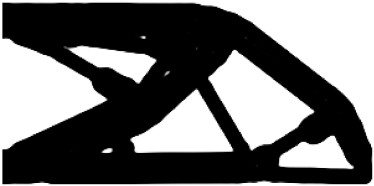}		
	&   \includegraphics[width=1.0\linewidth]{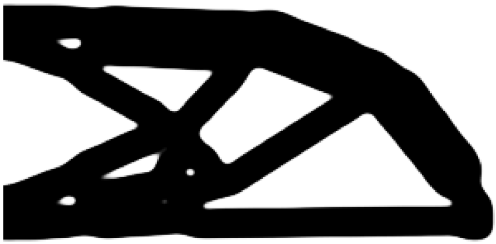}	
	&   \includegraphics[width=1.0\linewidth]{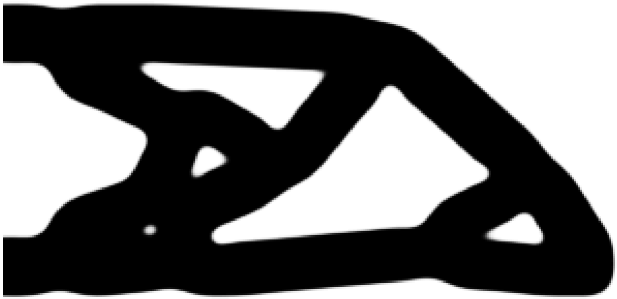}
	\\
	{\rotatebox{90}{{$\bm{\bar{\rho}}^\text{\tiny{AM}}$ , $\bm{\bar{\rho}}^\mathrm{ero}$}}}
	&   \includegraphics[width=1.0\linewidth]{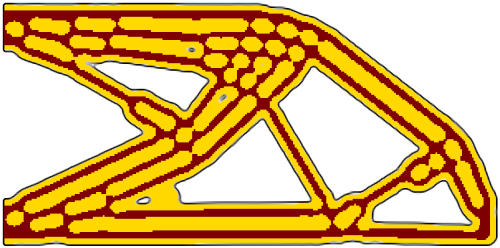}		
	&   \includegraphics[width=1.0\linewidth]{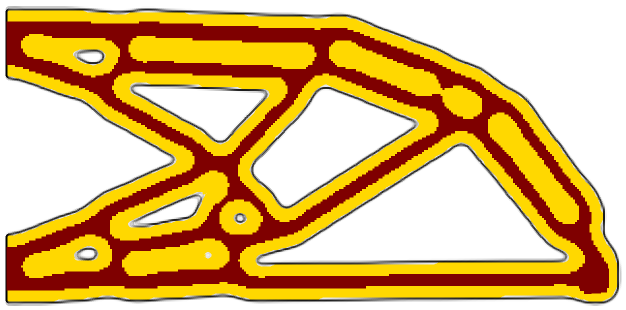}	
	&   \includegraphics[width=1.0\linewidth]{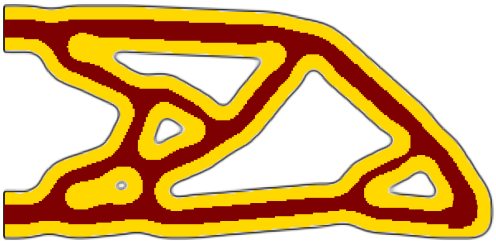}
	\\
	{\rotatebox{90}{{Depos. Path}}}
    &   \includegraphics[width=1.0\linewidth]{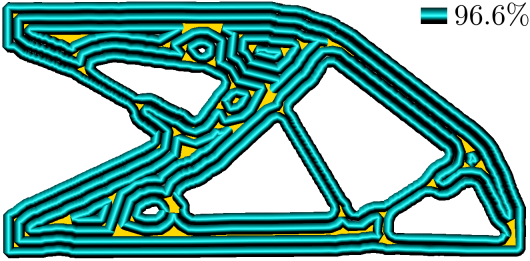}		
	&   \includegraphics[width=1.0\linewidth]{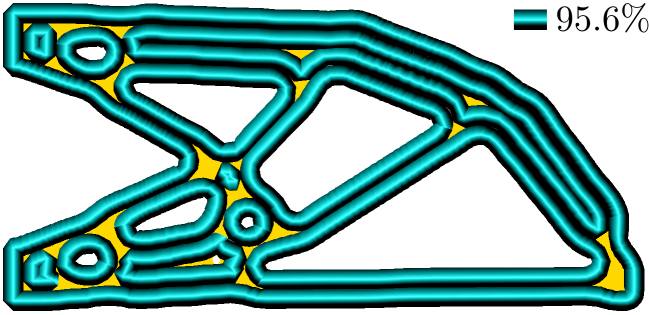}	
	&   \includegraphics[width=1.0\linewidth]{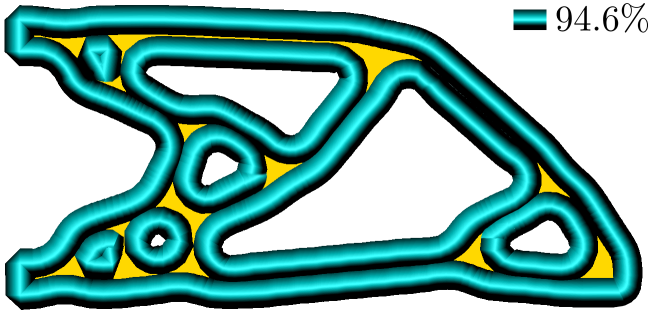}
	\\
	\bottomrule	
\end{tabular}
\label{TAB:MaxSize_2}	
\end{table}

Table \ref{TAB:MaxSize_2} shows three AM-constrained results obtained with varying nozzle sizes, and a volume fraction of $V^\text{\tiny{AM}} = 60 \%$. The Table includes the eroded designs superimposed on the optimized designs and the deposition path obtained with \linebreak PrusaSlicer. Note that the deposition paths are obtained with a nozzle size $r_\mathrm{nozzle}=0.5r_\mathrm{min.Solid}^\text{\tiny{AM}}$, which places at least 2 parallel paths per solid member. If the deposition nozzle were chosen twice the size, i.e.~$r_\mathrm{nozzle}=r_\mathrm{min.Solid}^\text{\tiny{AM}}$, then each solid member would consist of at least one bead. In such a case, the eroded and intermediate fields could be read as the deposition trajectory, since due to the projection strategy, they are positioned exactly at the center of the design intended for manufacturing, as shown in the second row of Table \ref{TAB:MaxSize_2}. Therefore, the deposition path provided by the eroded or intermediate design is expected to leave fewer unfilled zones than the one proposed by PrusaSlicer.

An interesting observation can be made from the results depicted in Table \ref{TAB:MaxSize_2}, where the AM-constrained designs present material beads, at first sight, oriented with the major principal stress direction. Surely, this pattern is due to the compliance cost function that is evaluated for a single load case, yet, it might be beneficial when considering an orthotropic elasticity of the material bead, for instance, in the WAAM process \citep{Laghi2020}. This observation, however, is not addressed in this article, hence it is left open for future research.

As the design intended for manufacturing corresponds to a dilated one, it is not possible to guarantee the minimum size of the cavities in $\bm{\bar{\rho}}^\text{\tiny{AM}}$. In addition, the presence of small cavities can be a consequence of utilizing large material beads. Note that in practice it is difficult to achieve high accuracy in the material bead width for some LSAM processes, as for instance WAAM \citep{xia2020}. In such instances, some dimensional uncertainty is present that could fill the voids present in the design. Such beneficial uncertainty could potentially be considered in the topology optimization formulation by using perturbation techniques on the length scale \citep{Lazarov2012}. This topic, however, is outside the scope of this work.

The approach that defines Method 2 is based on constraining the maximum size of a design and dilating it according to the prescribed deposition nozzle size. This results in a dilated design with improved printability, and in an eroded design that is placed at the center of the dilated members resembling a structural skeleton that can be interpreted as the deposition path. It should be noted that the structural skeleton is a by-product of constraining the intermediate design, and is not explicitly computed using edge detection algorithms \citep{Zhou2015}. This is the main advantage of the proposed Method 2 in comparison to other methods that could eventually be used for improving the printability of the designs, since, to date, the edge detection algorithms for topology optimization are not robust in the density method, mainly due to the fact that intermediate densities complicate boundary detection. For this reason, in the density method, length scale control algorithms based on the explicit calculation of the structural skeleton are applied almost at the end of the optimization process \citep{Zhou2015}, when the topology has already been defined, i.e.~they operate as post-processing tools. That being said, \citet{Liu2019} proposes a functional for level set topology optimization that allows to define a variable length scale within the design, which would also allow generation of designs adapted to the deposition nozzle, similar to Method 2. However, the cited method has not been validated in the density framework nor in 3D design domains, these two aspects being a major challenge for algorithms based on the explicit calculation of the skeleton. The following section demonstrates that the proposed Method 2 can be easily applied to 3D geometries, either considering AM processes with a fixed printing direction or with variable printing direction, such as AM processes assisted by robotic arms.

\section{3D example and discussion} \label{sec:5}

\begin{figure}[t!]
	\centering
    \includegraphics[width=0.7\linewidth]{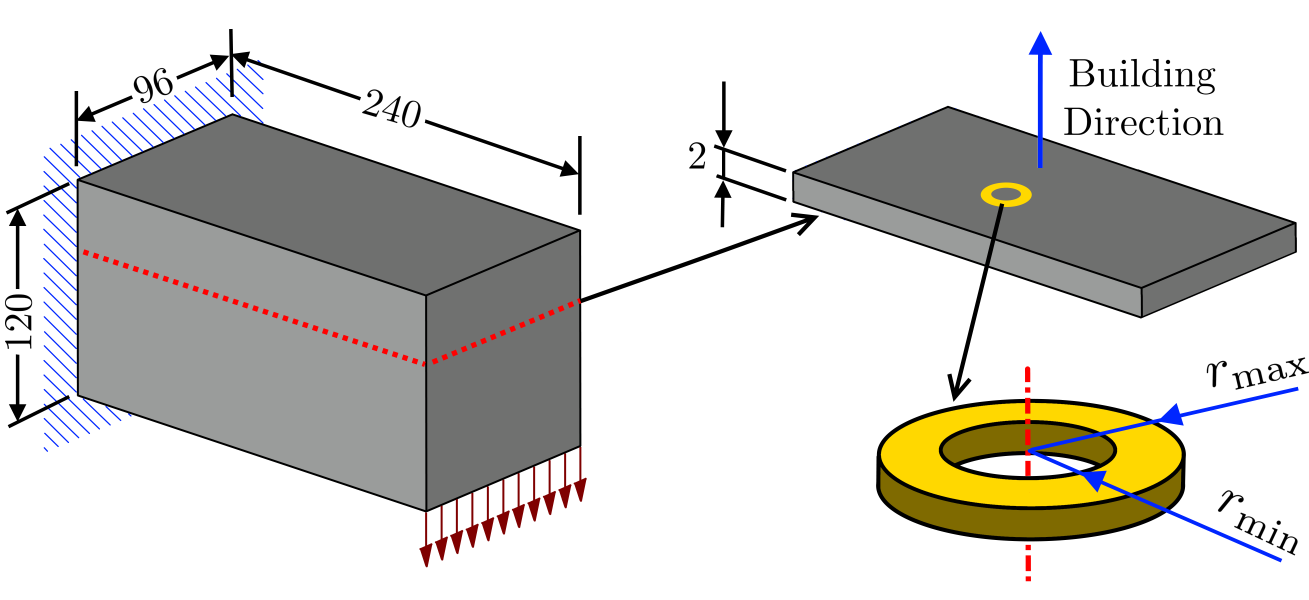}
    \vspace{-3mm}\\
    \subfigure[]{\label{FIG:3D_Des_a}}
	\hspace{50mm}~	
	\subfigure[]{\label{FIG:3D_Des_b}	}
	\\	
	\caption{In (a), the 3D cantilever beam for compliance minimization. In (b), the layer and the local region $\Omega_i$ where the local maximum size restriction is formulated. The design domain is discretized with $240 \times 120 \times 96$ finite elements.}	
	\label{FIG:3D_Des}			
\end{figure}

The two methods proposed in the previous sections were presented using a two-dimensional design case, where the deposition path is defined in a 2D plane. This section assesses and compares the methods using a three-dimensional design case. To this end, the cantilever beam shown in Fig.~\ref{FIG:3D_Des_a} is used. The design domain is discretized into 2.76 million cubic elements of 1 mm edge length. The optimized design is intended for an AM process equipped with a deposition nozzle of $r_\mathrm{nozzle}=5$ mm. It is assumed that the AM process generates layers orthogonal to the build direction, as shown in Fig.~\ref{FIG:3D_Des_b}. In addition, it is assumed that each layer is 2 mm thick, so the final component is made up of 60 layers. As each layer of the optimized design must conform to the size of the deposition nozzle, the maximum size restriction is applied in the plane, as shown in Fig.~\ref{FIG:3D_Des_b}.

As in the previous section, the finite element model considers an isotropic linear elastic behavior for both the deposited material and interlayer bond. Undoubtedly, this consideration may not be representative of some materials and LSAM processes, such as stainless steel in WAAM \citep{Kyvelou2020}, where material anisotropy should not be disregarded during topology optimization \citep{Zhang2017b}. However, as this work focuses on the nozzle size constraint, we keep the material modeling part general assuming an isotropic behavior instead of using a specific anisotropic model \citep[see, for instance,][]{Jantos2020}.

\begin{table}[t!]
\caption{3D cantilever beams obtained from the Reference, the Maximum-size-constrained and the AM-constrained topology optimization problems. The design domain is shown in Fig.~\ref{FIG:3D_Des}. The optimized designs are sliced using PrusaSlicer (v.2.2.0). The percentage of the design that is filled with deposited material is indicated next to the sliced design. The second and third rows show the least and most filled layers, respectively. The fourth row shows two representative layers for the deposition paths obtained with PrusaSlicer.}
\centering
\begin{tabular}{|m{0.3cm} |m{4.12cm}| m{4.12cm}| m{4.12cm}|}
	\toprule	    
    & \hspace{10mm}Reference & \hspace{0mm} Max.-Size-Constrained & \hspace{5mm}AM-Constrained  
	\\
	\cmidrule(r){1-4}
	{\rotatebox{90}{{Sliced Design}}}
	&   \includegraphics[width=1.0\linewidth]{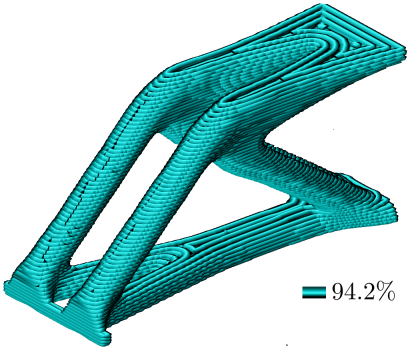}		
	&   \includegraphics[width=1.0\linewidth]{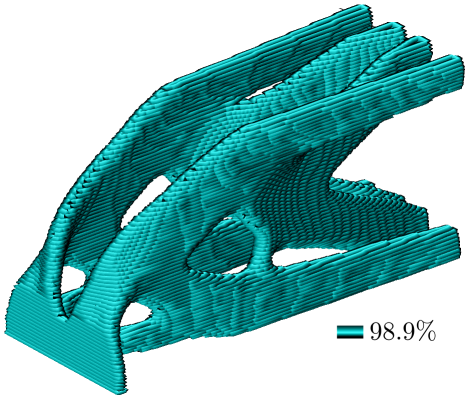}	
	&   \includegraphics[width=1.0\linewidth]{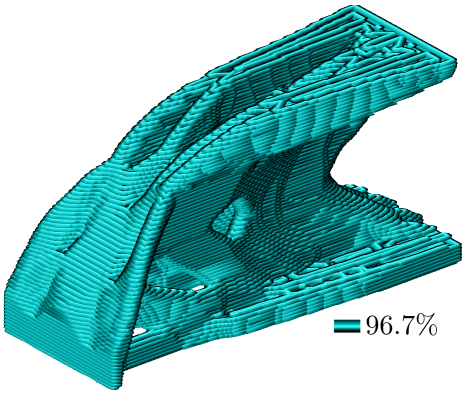}
	\\
	 & \hspace{8mm} \small{c($\bm{\bar{\rho}}^\mathrm{int}$)=1.00}
	 & \hspace{8mm} \small{c($\bm{\bar{\rho}}^\mathrm{int}$)=1.21}
	 & \hspace{8mm} \small{c($\bm{\bar{\rho}}^\text{\tiny{AM}}$)=1.16}
	\\
	\cmidrule(r){1-4}
	{\rotatebox{90}{{Least filled}}}
    &   \includegraphics[width=1.0\linewidth]{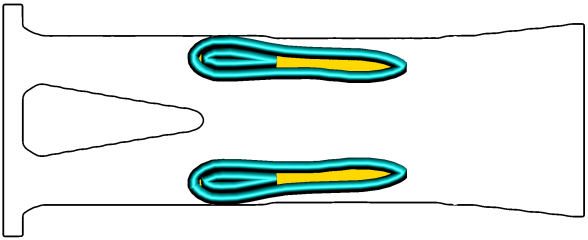}		
	&   \includegraphics[width=1.0\linewidth]{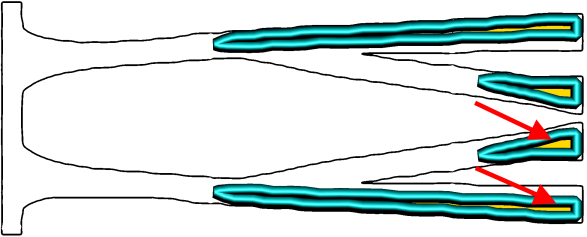}	
	&   \includegraphics[width=1.0\linewidth]{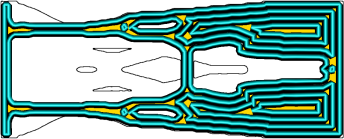}
	\\
	& \hspace{2mm} \small{Layer: 45 \;\;,\;  \scalerel*{\includegraphics{Figures/Printability.png}}{B} 86.6$\%$}
	& \hspace{2mm} \small{Layer: 57 \;\;,\;  \scalerel*{\includegraphics{Figures/Printability.png}}{B} 96.6$\%$} 
	& \hspace{2mm} \small{Layer: 2 \;\;,\;  \scalerel*{\includegraphics{Figures/Printability.png}}{B} 95.2$\%$}
	\\
	\cmidrule(r){1-4}
	{\rotatebox{90}{{Most filled}}}
    &   \includegraphics[width=1.0\linewidth]{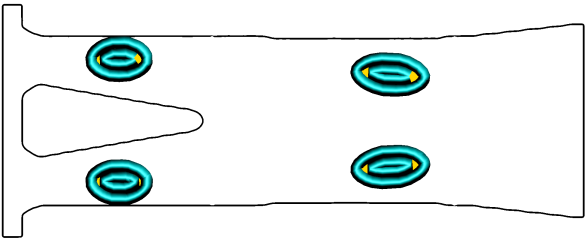}		
	&   \includegraphics[width=1.0\linewidth]{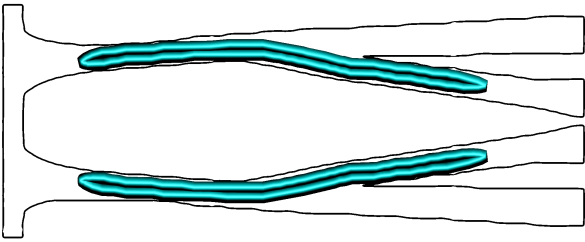}	
	&   \includegraphics[width=1.0\linewidth]{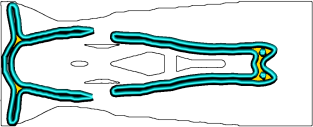}
	\\
	& \hspace{1mm} \small{Layer: 22 \;\;,\;  \scalerel*{\includegraphics{Figures/Printability.png}}{B} 97.6$\%$ }
	& \hspace{1mm} \small{Layer: 42 \;\;,\;  \scalerel*{\includegraphics{Figures/Printability.png}}{B} 100.0$\%$ }
	& \hspace{1mm} \small{Layer: 15 \;\;,\;  \scalerel*{\includegraphics{Figures/Printability.png}}{B} 97.5$\%$}
	\\
	\cmidrule(r){1-4} 
	\multirow{3}{*}[2.0ex]{\rotatebox{90}{Representative Layers}}
    &   \includegraphics[width=1.0\linewidth]{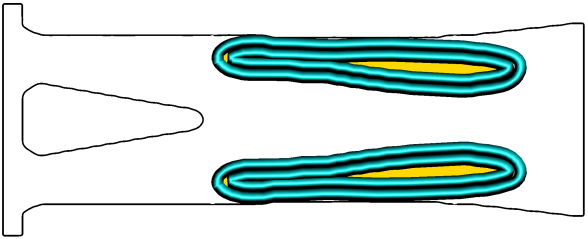}		
	&   \includegraphics[width=1.0\linewidth]{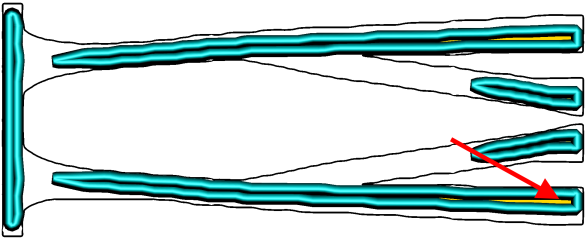}	
	&   \includegraphics[width=1.0\linewidth]{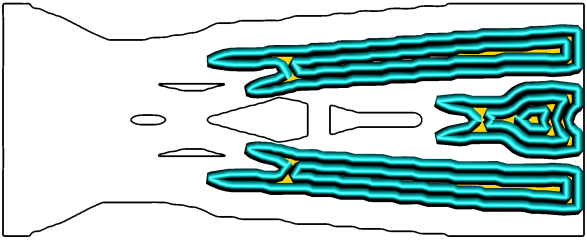}
	\\
	& \hspace{1mm} \small{Layer: 53 \;\;,\;  \scalerel*{\includegraphics{Figures/Printability.png}}{B} 93.0$\%$ }
	& \hspace{1mm} \small{Layer: 6 \;\;,\;  \scalerel*{\includegraphics{Figures/Printability.png}}{B} 98.7$\%$ }
	& \hspace{1mm} \small{Layer: 42 \;\;,\;  \scalerel*{\includegraphics{Figures/Printability.png}}{B} 97.4$\%$}
	\\
	&   \includegraphics[width=1.0\linewidth]{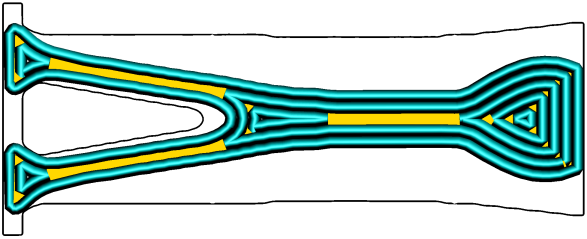}		
	&   \includegraphics[width=1.0\linewidth]{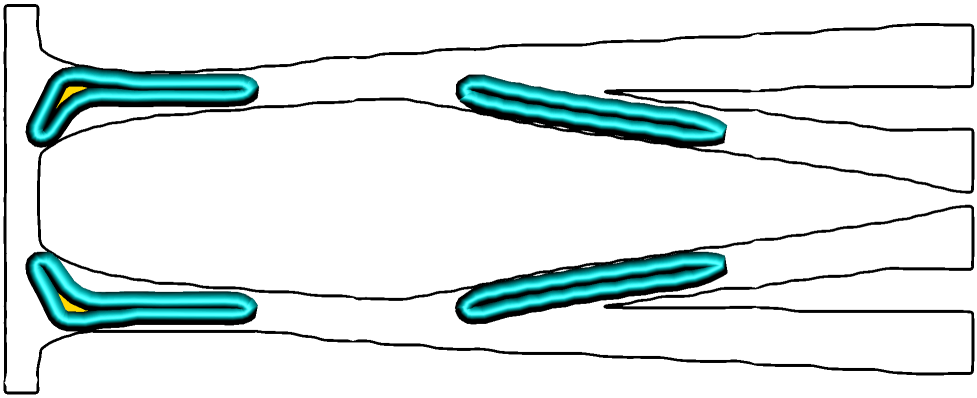}	
	&   \includegraphics[width=1.0\linewidth]{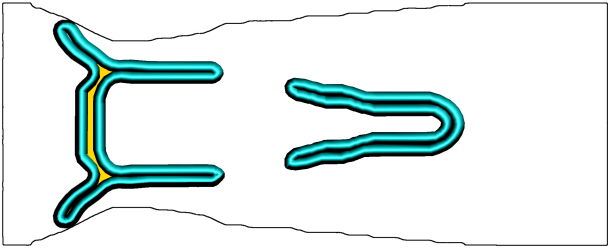}
	\\
	& \hspace{1mm} \small{Layer: 6 \;\;,\;  \scalerel*{\includegraphics{Figures/Printability.png}}{B} 89.5$\%$ }
	& \hspace{1mm} \small{Layer: 24 \;\;,\;  \scalerel*{\includegraphics{Figures/Printability.png}}{B} 99.1$\%$} 
	& \hspace{1mm} \small{Layer: 30 \;\;,\;  \scalerel*{\includegraphics{Figures/Printability.png}}{B} 97.0$\%$}
	\\
	\bottomrule	
\end{tabular}
\label{TAB:Results_3D_1}	
\end{table}

Three designs are reported, the Reference, the Maximum-size-constrained and the AM-constrained designs. The optimization problems are solved for a volume restriction of $15\%$ and with a minimum size of $r_\mathrm{min.solid} = 5$ mm. The optimized results are sliced in PrusaSlicer, and the obtained deposition paths are shown in Table \ref{TAB:Results_3D_1}. The first row of the table shows the component after all the layers have been deposited and the averaged printability of all layers combined. The remaining rows show 4 selected layers and the associated deposition path. These layers correspond to the least filled, the most filled, and two representative of the average. The layers are shown inside the projected perimeter of the complete structure to provide guidance on the location of the layer with respect to the complete design.

From the printability indices of Table \ref{TAB:Results_3D_1}, it is clear that as in the 2D examples, the maximum-size-constrained design contains the least amount of unfilled zones. Also, since two material beads are placed per solid member width, no T-joints or path crossovers are obtained. In addition, due to the uniform size of the solid members, the size-constrained component resembles a design composed of walls of uniform width. As a number of AM processes are proficient in the production of walls, such as WAAM, it is expected that the size-restricted component will be the easiest to manufacture among the three reported designs.

The small unfilled gaps (1.1$\%$) in the maximum-size-constrained design are mainly zones where the local maximum-size constraint is violated, as those pointed out by arrows in Table \ref{TAB:Results_3D_1}. This drawback caused by the smoothed aggregation function could be reduced, but not completely eliminated, by using a finer mesh or larger aggregation exponents \citep{Fernandez2019}, i.e.~at the cost of increasing computational expense. Another approach to improve printability is to post-process the optimized designs using the geometric constraints proposed by \citet{Zhou2015}, which, in our experience, enforce length scale more strictly.  

Although the size-constrained design is the best candidate for manufacturing, it exhibits the lowest performance, since the maximum size restriction prevents bulky zones that are required to increase the structural stiffness. This could be even more detrimental in applications involving stress or fatigue constraints \citep{Collet2017}. Therefore, it is expected that for cases involving mechanical constraints, the AM-constrained problem will be a better choice for yielding high performance designs suitable for AM processes equipped with large deposition nozzles.

It was observed in the 2D examples that the eroded and intermediate designs can be interpreted as the structural skeleton or the deposition paths in the AM-constrained formulation. To illustrate this in the 3D design case, Table \ref{TAB:Results_3D_2} shows the AM-design and 3 representative layers of the component. Each layer displays the eroded design in red and the AM layer in yellow. The table contains 3 optimized solutions obtained with the same nozzle size but with different volume restriction.

\begin{table}
\caption{AM-constrained designs. The layers show the eroded design in red and the AM design in yellow.}
\centering
\begin{tabular}{|m{0.3cm} |m{4.12cm}| m{4.12cm}| m{4.12cm}|}
	\toprule	    
     & \hspace{10mm}$V^\text{\tiny{AM}} = 15 \%$ & \hspace{10mm} $V^\text{\tiny{AM}} = 20 \%$ & \hspace{10mm}$V^\text{\tiny{AM}} = 30 \%$  	
	\\
	\cmidrule(r){1-4}
	{\rotatebox{90}{{$\bm{\bar{\rho}}^\text{\tiny{AM}}$}}}
    &  \hspace{1mm} \includegraphics[width=0.95\linewidth]{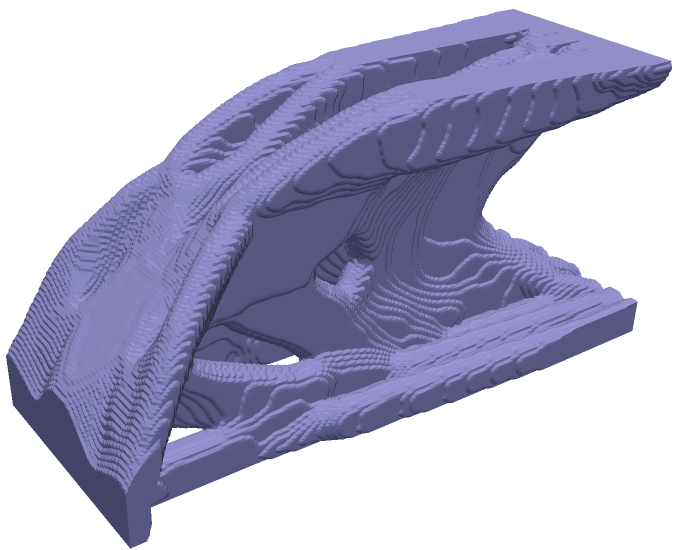}		
	&  \hspace{1mm} \includegraphics[width=0.95\linewidth]{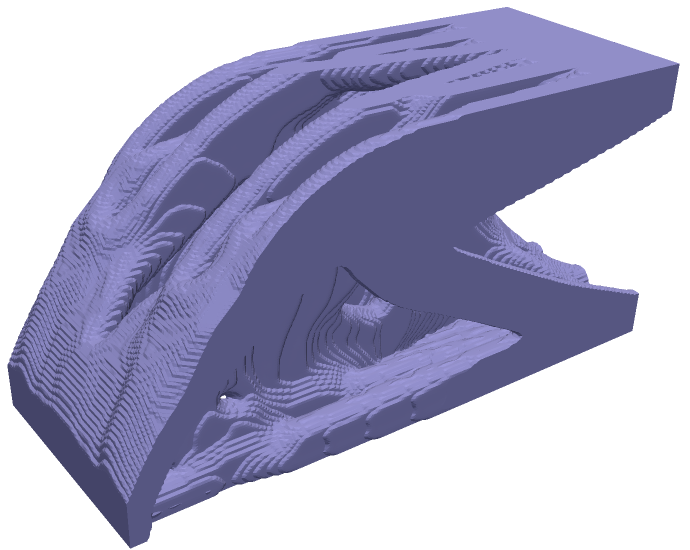}	
	&  \hspace{1mm} \includegraphics[width=0.95\linewidth]{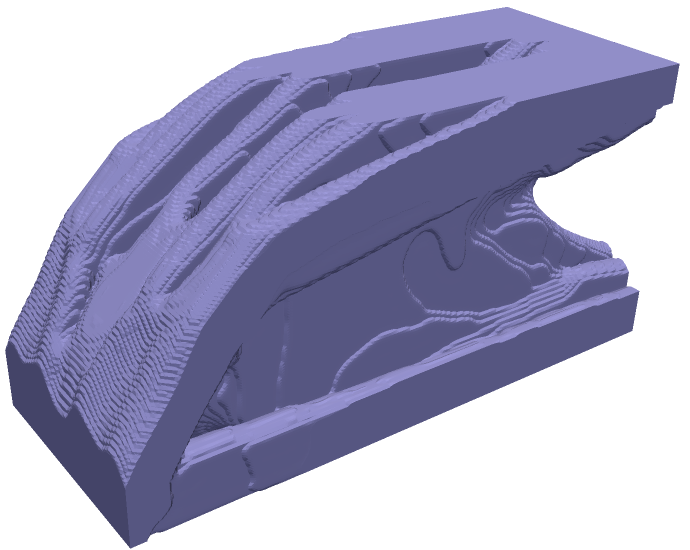}
	\\
	\cmidrule(r){1-4}
	{\rotatebox{90}{{Layer 2}}}
    & \vspace{2mm}\hspace{1mm}  \includegraphics[width=0.9\linewidth]{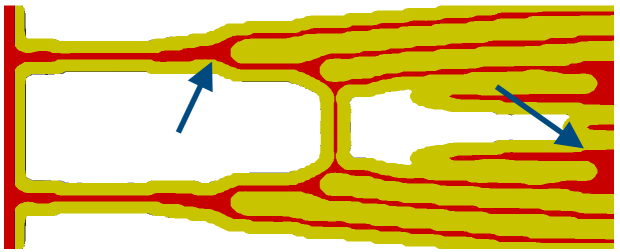}		
	& \vspace{2mm}\hspace{1mm}  \includegraphics[width=0.9\linewidth]{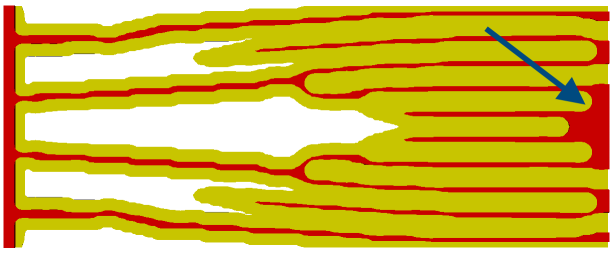}	
	& \vspace{2mm}\hspace{1mm}  \includegraphics[width=0.9\linewidth]{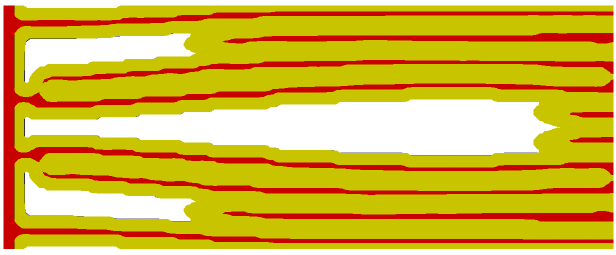}
	\\
	{\rotatebox{90}{{Layer 20}}}
    & \vspace{2mm}\hspace{1mm}  \includegraphics[width=0.92\linewidth]{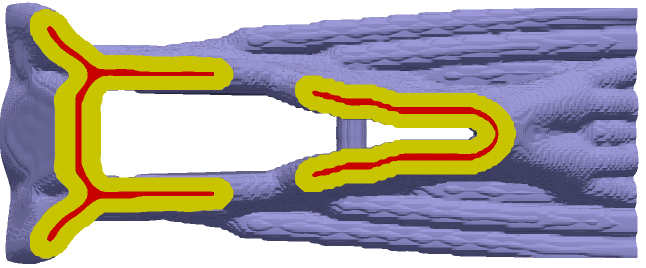}		
	& \vspace{2mm}\hspace{1mm}  \includegraphics[width=0.92\linewidth]{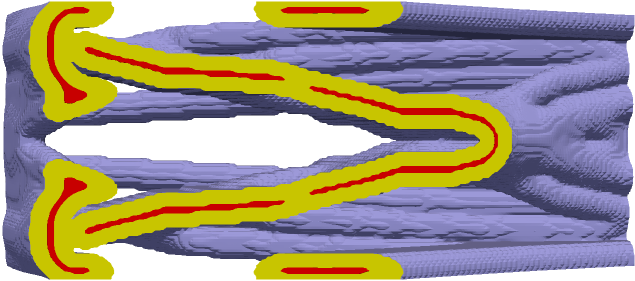}	
	& \vspace{2mm}\hspace{1mm}  \includegraphics[width=0.92\linewidth]{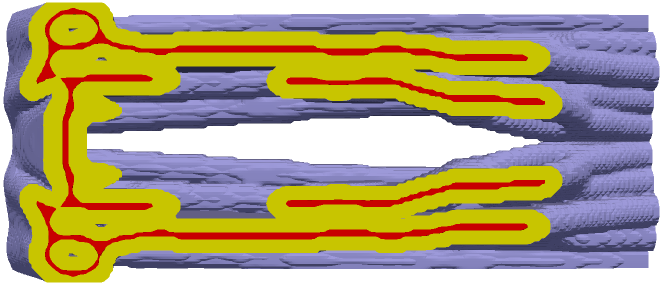}
	\\
	{\rotatebox{90}{{Layer 40}}}
    & \vspace{2mm}\hspace{1mm}  \includegraphics[width=0.93\linewidth]{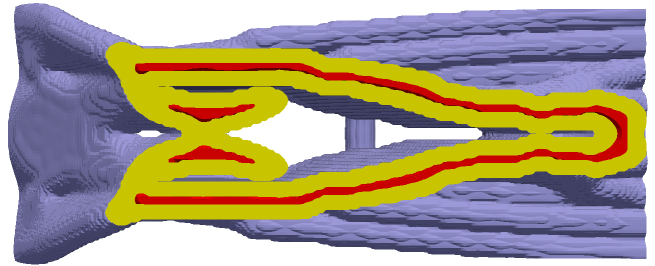}		
	& \vspace{2mm}\hspace{1mm}  \includegraphics[width=0.96\linewidth]{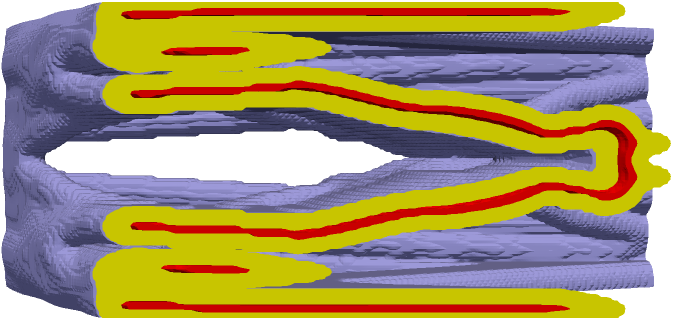}	
	& \vspace{2mm}\hspace{1mm}  \includegraphics[width=0.96\linewidth]{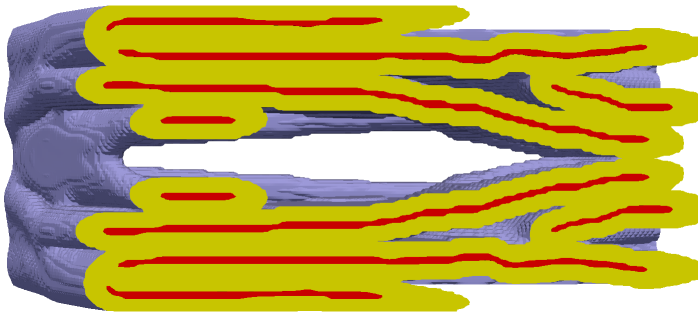}
	\\
	\bottomrule	
\end{tabular}
\label{TAB:Results_3D_2}	
\end{table}

From the eroded design it is rather intuitive to conceive a deposition path for the layer shown in Table \ref{TAB:Results_3D_2}. However, there are some places that could hinder the path interpretation, as those pointed out by arrows in the Layer 2 depicted in Table \ref{TAB:Results_3D_2}. These places are in violation of the local volume restriction that imposes the maximum size and, therefore, it is not possible to ensure that these zones will be filled during the LSAM process. Due to these zones in conflict with the local maximmum size restriction, the optimized design will likely need to be post-processed to conform these details to the resolution of LSAM process, for instance, using more strict geometric constraints on the optimized designs \citep{Zhou2015}.

\begin{table}
\caption{Design details of the 3D cantilever beams shown in Fig.~\ref{FIG:3D_Des_B}.}
\centering
\begin{tabular}{|m{0.3cm} | m{6.2cm}|  m{6.2cm}|}
	\toprule	    
     & \hspace{10mm}Maximum-Size-Constrained & \hspace{15mm}AM-Constrained  	
	\\
	\cmidrule(r){1-3}
	{\rotatebox{90}{{Optimized Design}}}
	&  \hspace{10mm} \includegraphics[width=0.75\linewidth]{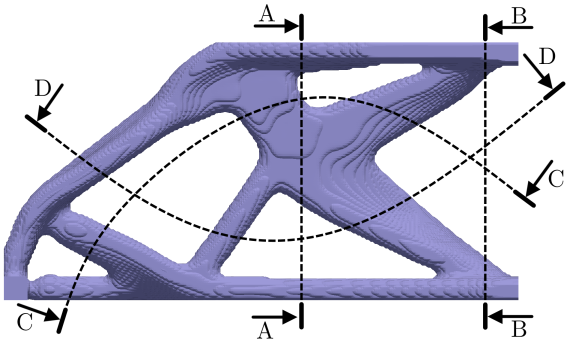}	
	&  \hspace{10mm} \includegraphics[width=0.75\linewidth]{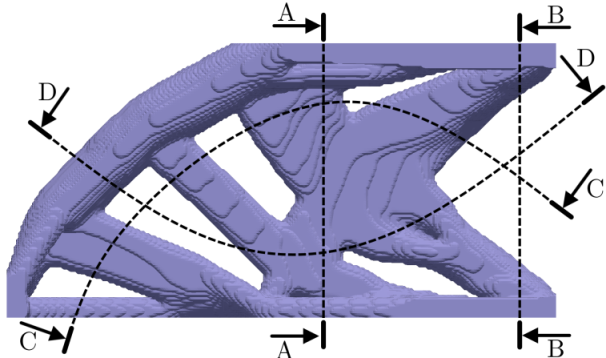}
	\\
	 & \hspace{28mm} c($\bm{\bar{\rho}}^\mathrm{int}$)=1.15
	 & \hspace{28mm} c($\bm{\bar{\rho}}^\text{\tiny{AM}}$)=1.08
	\\
	\cmidrule(r){1-3}
	{\rotatebox{90}{$\bm{\bar{\rho}}^\mathrm{ero}$}}
	&  \hspace{10mm} \includegraphics[width=0.78\linewidth]{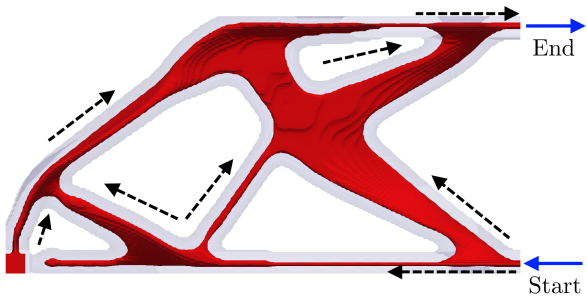}	
	&   \includegraphics[width=1.0\linewidth]{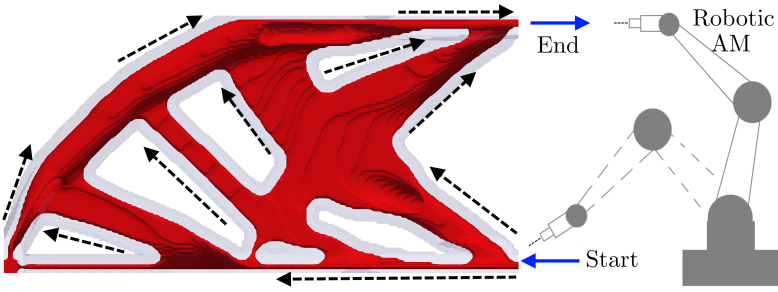}
	\\
	\cmidrule(r){1-3}
	{\rotatebox{90}{{Section A-A}}}
	&  \hspace{25mm} \includegraphics[width=0.28\linewidth]{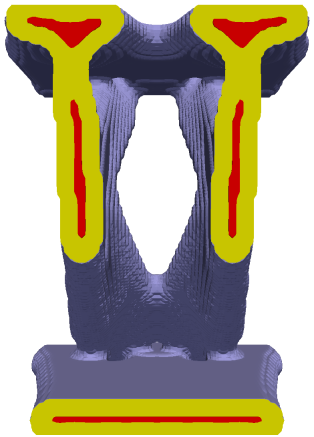}	
	&  \hspace{25mm} \includegraphics[width=0.32\linewidth]{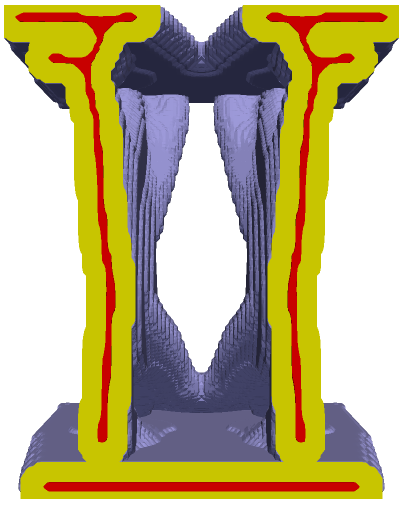}
	\\
	\cmidrule(r){1-3}
	{\rotatebox{90}{{Section B-B}}}
	&  \vspace{1mm}\hspace{25mm} \includegraphics[width=0.28\linewidth]{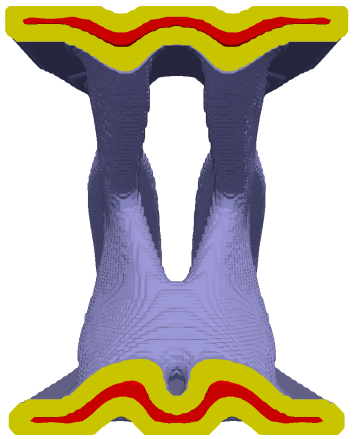}	
	&  \vspace{1mm}\hspace{26mm} \includegraphics[width=0.28\linewidth]{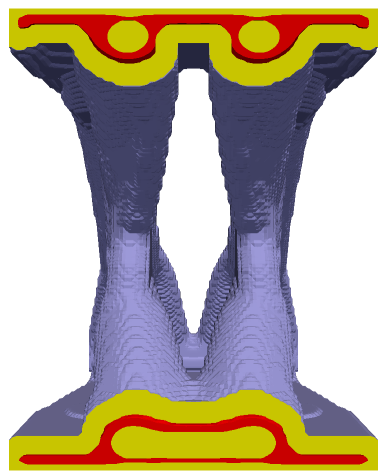}
	\\
	\cmidrule(r){1-3}
	{\rotatebox{90}{{Section C-C}}}
	&  \hspace{12mm} \includegraphics[width=0.67\linewidth]{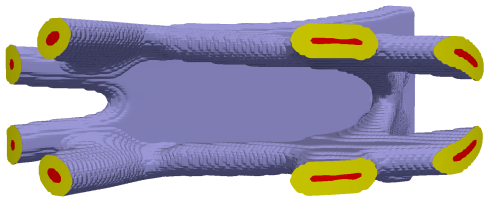}	
	&  \hspace{12mm} \includegraphics[width=0.67\linewidth]{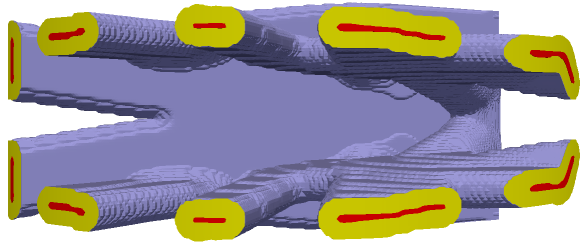}
	\\
	\cmidrule(r){1-3}
	{\rotatebox{90}{{Section D-D}}}
	&  \hspace{12mm} \includegraphics[width=0.67\linewidth]{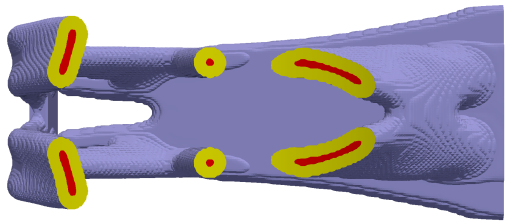}	
	&  \hspace{12mm} \includegraphics[width=0.67\linewidth]{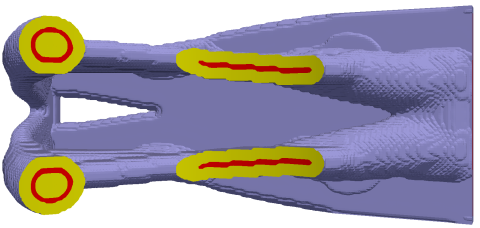}
	\\
	\bottomrule	
\end{tabular}
\label{TAB:Results_3D_3}	
\end{table}

\begin{figure}
	\centering
    \subfigure[Size-Constrained]{
    		\includegraphics[width=0.36\linewidth]{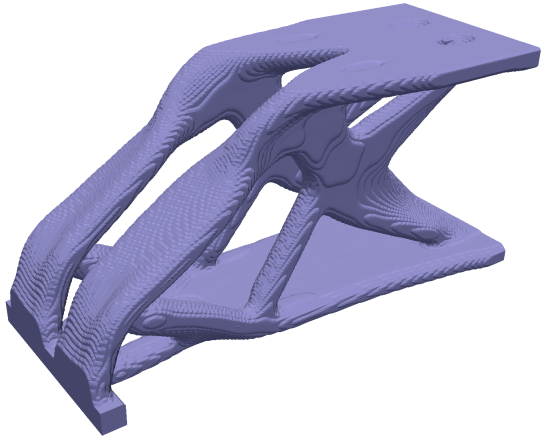}			
		\label{FIG:3D_Des_B_a}	
	}	
	\subfigure[AM-Constrained]{
		    \includegraphics[width=0.36\linewidth]{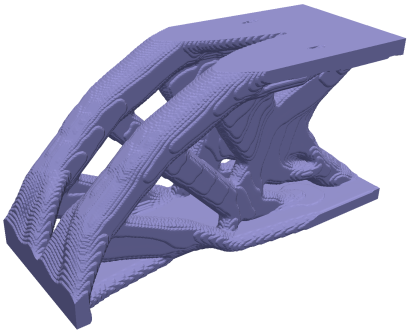}	
			\label{FIG:3D_Des_B_b}	
	}
	\\
	\caption{Optimized 3D cantilever beam for compliance minimization.}	
	\label{FIG:3D_Des_B}			
\end{figure}

Notably, the 3D examples shown in this section assume a fixed printing direction. Nevertheless, there are Large-Scale AM processes that are able to change the stacking direction of the layers, either by the rotation of the baseplate or of the deposition tool. Interestingly, the methods proposed in this paper can be easily adapted to such processes. For this purpose, the local $\Omega_i$ region where the maximum size restriction is evaluated should be defined as a sphere instead of a two-dimensional disk, as presented in \citep{Fernandez2020}. For instance, Fig.~\ref{FIG:3D_Des_B} shows the size-constrained and AM-constrained designs obtained by using a spherical $\Omega_i$ region. The volume constraint is set as $V = 15 \%$ and the length scale is defined identical to the previous examples. The inner details of these designs are summarized in Table \ref{TAB:Results_3D_3}. The first row of the table contains the optimized designs and the sections planes that display the working layer. The second row shows the eroded design and the printing direction devised intuitively. The remaining rows display the sections, where the eroded design is shown in red and the cross-section in yellow.	

The designs in Table \ref{TAB:Results_3D_3} are considerably different from their counterparts obtained with a fixed vertical printing direction (Table \ref{TAB:Results_3D_1}). The designs having the printing direction freedom possess two thick bars at the loaded side (see also Figure \ref{FIG:3D_Des_B}) that connect the upper part of the design with the load transmission zone. These bars are also present in the reference design (Table \ref{TAB:Results_3D_1}), which would explain the high performance of these solutions (Table \ref{TAB:Results_3D_3}) with respect to those obtained using a fixed building orientation (Table \ref{TAB:Results_3D_1}). Interestingly, these two curved bars present in the designs of Table \ref{TAB:Results_3D_3} possess different geometry depending on the formulation used. As can be seen in section D-D, these bars are flat in the size-constrained design, while they are round in the AM-constrained design. Although both geometries are suited to the size of the deposition nozzle, the round bar offers a higher area moment of inertia and hence a higher bending stiffness.

As before with the fixed printing direction, the AM-constrained design shows better performance than the maximum-size constrained one, presumably because it allows thicker members where this is beneficial.

\section{Other test cases and TO problems} \label{sec:5.5}

The proposed Method 2, based on dilating a design constrained in maximum size, was introduced in Eq.~\eqref{eq:Reference_Opti_AM} using the compliance minimization problem subject to a volume constraint. This formulation was chosen for the sake of simplicity, but in fact, Method 2 is applicable to any topology optimization problem where the maximum size constraint (Eq.~\eqref{eq:GMS}) has been validated. Moreover, in our experience, the maximum size constraint, and hence Method 2, can be applied in most of fields where the robust formulation based on the eroded, intermediate and dilated designs has been implemented. For the sake of demonstration, in this section we therefore present results of other topology optimization problems. 

In the following, the proposed Method 2 is considered under compliance minimization subject to a volume and stress constraints. The topology optimization problem can be written as:
\begin{align} \label{eq:Reference_Opti_AM_Stress}
	\begin{split}
  		{\min_{\rho}} &\quad  \alpha \: c(\bm{\bar{\rho}}^{\text{\tiny{AM}}}) + (1-\alpha) \: c(\bm{\bar{\rho}}^{\mathrm{ero}}) 			\\
	  	\mathrm{s.t. :} &\quad \mathbf{v}^{\intercal} \bm{\bar{\rho}}^{\text{\tiny{AM}}} \leq V^{\text{\tiny{AM}}} 	\\
	  			& \quad \mathrm{G_{ms}}( \bm{\bar{\rho}}^{\mathrm{int}}) \leq 0 	\\
	  			& \quad \mathrm{max}( \bm{\sigma} (\bm{\bar{\rho}}^{\text{\tiny{AM}}})) \leq \sigma^* 	\\
	  			&\quad 0 \leq {\rho_i} \leq1 \; ,
	\end{split}
\end{align}
\noindent where $\bm{\sigma} (\bm{\bar{\rho}}^{\text{\tiny{AM}}})$ represents the equivalent Von Mises stress computed in the dilated field and $\sigma^*$ is the user-defined stress limit. The max($\cdot$) function is computed using a \textit{p-mean} approximation, following exactly the same implementation as \citet{Verbart2017}. The equivalent Von Mises stress of the dilated field is computed as described in \citep{Da2019}. For the sake of brevity, we omit the numerous equations defining the stress constraint and interested readers are referred to the cited works.

The classical L-beam with a volume constraint of $40\%$ is considered. For this test case, three topology optimization problems are solved, the reference (Eq.~\ref{eq:Reference_Opti}), the AM-constrained (Eq.~\ref{eq:Reference_Opti_AM}), and the AM-stress-constrained one (Eq.~\ref{eq:Reference_Opti_AM_Stress}). In the latter, the imposed stress limit ($\sigma^*$) is 0.8 times the maximum stress computed in the reference design. The three optimized designs are shown in the first row of Table \ref{TAB:Results_LShape}.

\begin{table}[t!]
\caption{L-Shaped beam optimized for compliance minimization. 200 finite elements discretize the long edge of the beam. The minimum size of the solid phase is defined by a circle of 4 finite element radius. In these problems, the maximum Heaviside projection exponent ($\beta$) is set to 20, and the SIMP exponent is fixed at 3.0.}
\centering
\begin{tabular}{|m{0.25cm} |m{4.08cm}| m{4.08cm}| m{4.08cm}|}
	\toprule	    
    & \hspace{10mm}Reference & \hspace{0mm} AM-Constrained & \hspace{5mm}AM-Stress-Constr.  
	\\
	\cmidrule(r){1-4}
	{\rotatebox{90}{{Optimized Design}}}
	&   \includegraphics[width=0.84\linewidth]{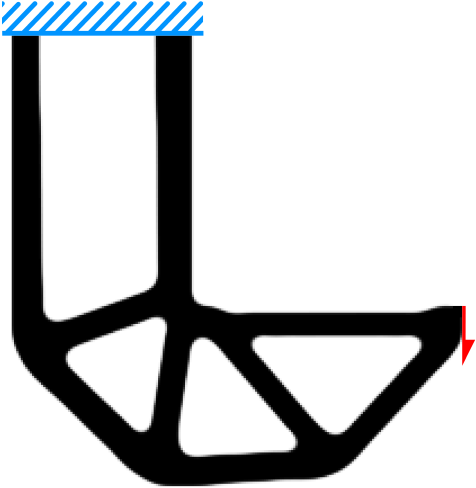}		
	&   \includegraphics[width=0.84\linewidth]{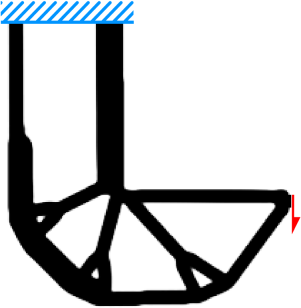}	
	&   \includegraphics[width=0.84\linewidth]{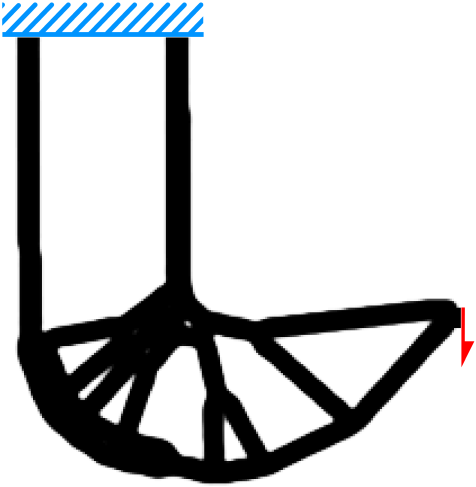}
	\\
	 & \hspace{-2mm} \small{c(}\footnotesize{$\bm{\bar{\rho}}^\mathrm{int}$}\small{)=1.00, max($\bm{\sigma}$)=1.00}
	 & \hspace{-2.5mm} \small{c(}\footnotesize{$\bm{\bar{\rho}}^{\text{\tiny{AM}}}$}\small{)=1.09, max($\bm{\sigma}$)=1.01}
	 & \hspace{-2.5mm} \small{c(}\footnotesize{$\bm{\bar{\rho}}^{\text{\tiny{AM}}}$}\small{)=1.24, max($\bm{\sigma}$)=0.86}
	\\
	\cmidrule(r){1-4}
	{\rotatebox{90}{{Stress field $\bm{\sigma}$}}}
    &   \includegraphics[width=0.84\linewidth]{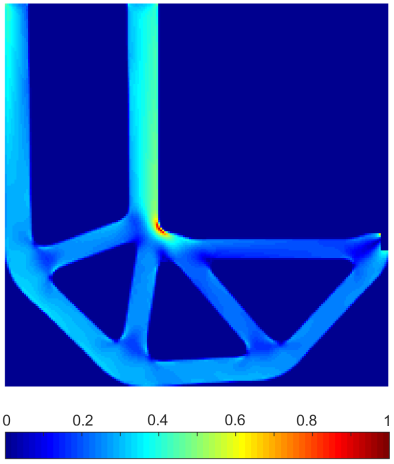}		
	&   \includegraphics[width=0.84\linewidth]{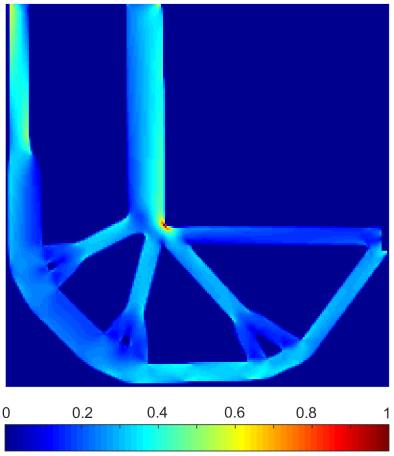}	
	&   \includegraphics[width=0.84\linewidth]{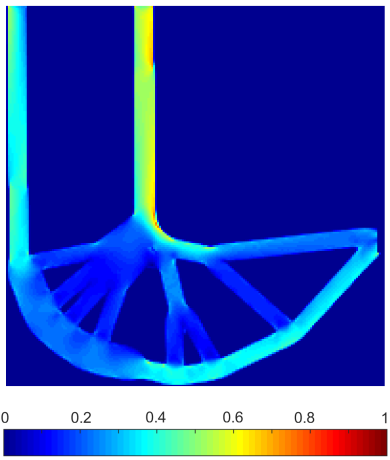}
	\\
	\cmidrule(r){1-4}
	{\rotatebox{90}{{$\bar{\bm{\rho}}^\mathrm{ero}$ and Design field}}}
    &   \includegraphics[width=0.84\linewidth]{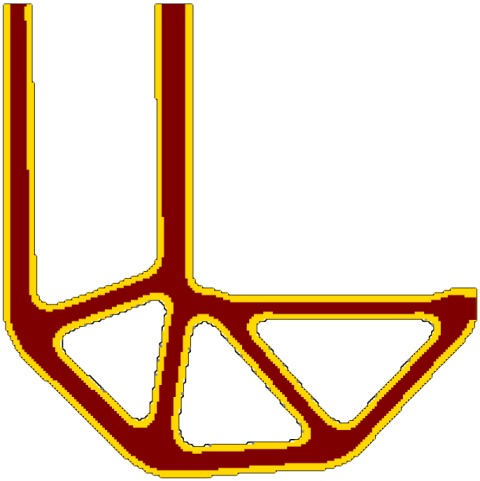}		
	&   \includegraphics[width=0.84\linewidth]{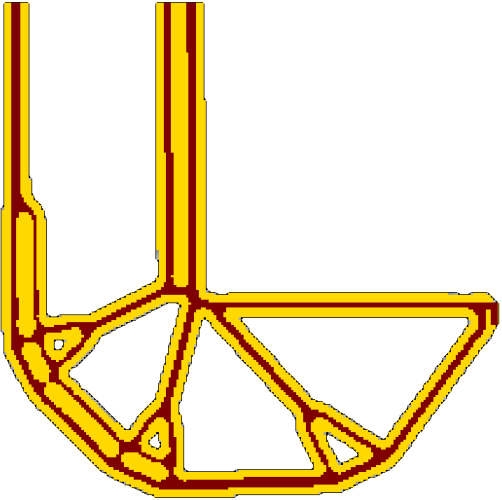}	
	&   \includegraphics[width=0.84\linewidth]{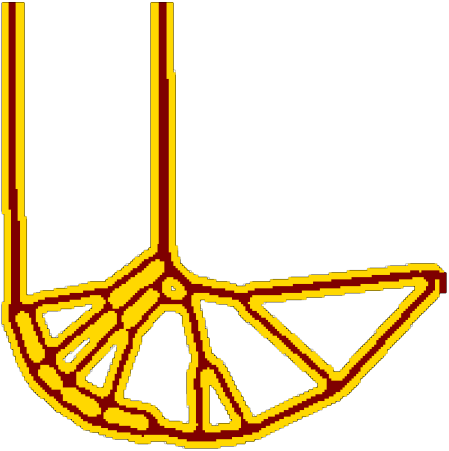}
	\\
	\cmidrule(r){1-4} 
	\multirow{3}{*}[5.0ex]{\rotatebox{90}{Perimeter path}}
    &   \includegraphics[width=0.84\linewidth]{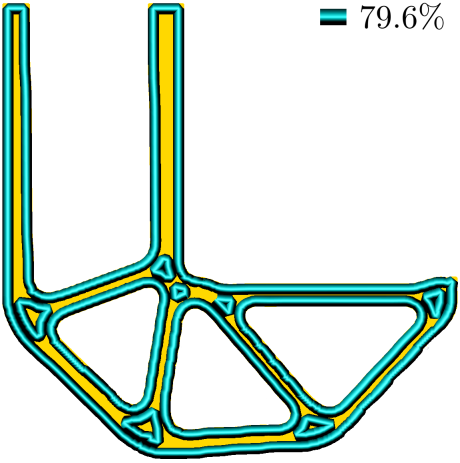}		
	&   \includegraphics[width=0.84\linewidth]{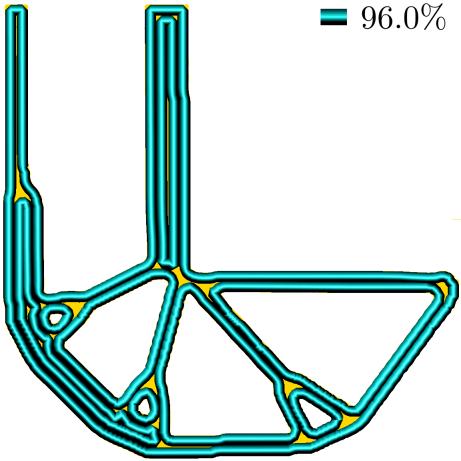}	
	&   \includegraphics[width=0.84\linewidth]{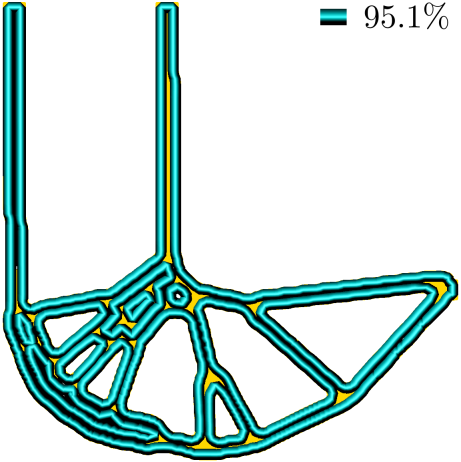}
	\\
	\bottomrule	
\end{tabular}
\label{TAB:Results_LShape}	
\end{table}

The two AM-constrained results in Table \ref{TAB:Results_LShape} (second and third columns) clearly reach the intended purpose, which is the production of designs adapted to the deposition nozzle size. In addition, in such designs, the eroded field can be interpreted as the deposition path, as shown in the third row of Table \ref{TAB:Results_LShape}. However, the introduction of stress constraints produces a different arrangement of material beads. The stress-constrained design places the material beads mostly around the vicinity of the L-beam curvature, which, along with the smooth curvature of the design, reduces the maximum stresses in the structure compared to the other two designs. This solution demonstrates the feasibility of implementing the proposed Method 2 in problems involving stress constraints.

Although the combination of Method 2 with stress constraints is demonstrated, there are still aspects that can be improved. For example, it can be seen that the AM-stress-constrained design does not accurately meet the imposed stress constraint. This is probably due to the p-mean function, which underestimates the maximum stress value for the sake of a differentiable approximation, therefore other relaxation strategies should be examined \citep{Da2019}. Also, observations regarding small gaps in the deposition paths hold as before, which could result in potential stress concentrations. A more in-depth study of these aspects is identified as a direction for future research.

To close this section, we include results for the well-known force inverter benchmark test. The AM-constrained formulation of the force inverter can be written as follows:  
\begin{align} \label{eq:Reference_Opti_AM_Force_Inverter}
	\begin{split}
  		{\min_{\rho}} &\quad  \mathrm{max}(c(\bm{\bar{\rho}}^{\mathrm{ero}}),c(\bm{\bar{\rho}}^{\mathrm{int}})) \: (1-\alpha) + c(\bm{\bar{\rho}}^{\text{\tiny{AM}}}) \: \alpha  			\\
	  	\mathrm{s.t. :} &\quad \mathbf{v}^{\intercal} \bm{\bar{\rho}}^{\text{\tiny{AM}}} \leq V^{\text{\tiny{AM}}} 	\\
	  			& \quad \mathrm{G_{ms}}( \bm{\bar{\rho}}^{\mathrm{int}}) \leq 0 	\\
	  			&\quad 0 \leq {\rho_i} \leq1 \; ,
	\end{split}
\end{align}
\noindent where $c$ represents the output displacement of the corresponding design (considered negative due to the adopted reference system). The force inverter is modeled under linear elasticity, as is done in many other works in the literature \citep{Sigmund1997,Bendsoe2013}, so further formulation and implementation details are omitted for the sake of brevity and the reader is referred to the cited works.

\begin{table}
\caption{Half force inverter design solutions. Boundary conditions are illustrated in the reference design. There, the red arrow represents the external force and the blue arrow the output displacement to maximize. The orange dashed line is the symmetry line. The design domain is discretized using 200$\times$100 finite elements. The minimum size of the solid phase is defined by a circle of 4 finite element radius. $c$ represents the output displacement of the corresponding design. In these problems, the maximum Heaviside projection exponent ($\beta$) is set to 38, and the SIMP exponent is fixed at 3.0.}
\centering
\begin{tabular}{|m{0.3cm} |m{4.1cm}| m{4.1cm}| m{4.1cm}|}
	\toprule	    
    & \hspace{10mm}Reference & \hspace{0mm} Size-Constrained & \hspace{5mm}AM-Constrained.  
	\\
	\cmidrule(r){1-4}
	{\rotatebox{90}{\small{Design}}}
	&   \includegraphics[width=0.9\linewidth]{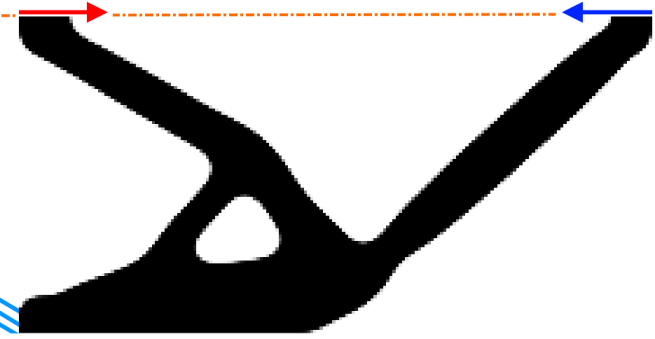}		
	&   \includegraphics[width=0.9\linewidth]{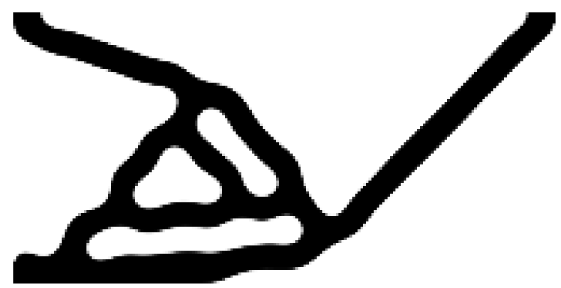}	
	&   \includegraphics[width=0.9\linewidth]{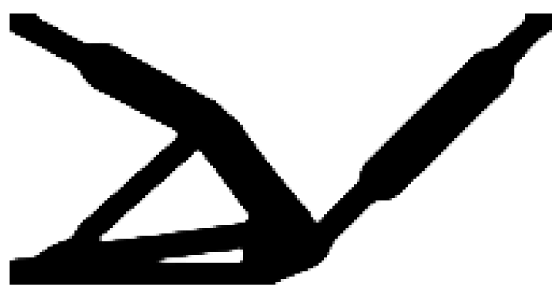}
	\\
	 & \hspace{7mm} \small{c(}\footnotesize{$\bm{\bar{\rho}}^\mathrm{int}$}\small{)= $-$1.00}
	 & \hspace{7mm} \small{c(}\footnotesize{$\bm{\bar{\rho}}^\mathrm{int}$}\small{)= $-$0.94}
	 & \hspace{7mm} \small{c(}\footnotesize{$\bm{\bar{\rho}}^{\text{\tiny{AM}}}$}\small{)= $-$0.97}
	\\
	\cmidrule(r){1-4}
	{\rotatebox{90}{\small{$\bar{\bm{\rho}}^\mathrm{ero}$ and Design}}}
    &   \includegraphics[width=0.9\linewidth]{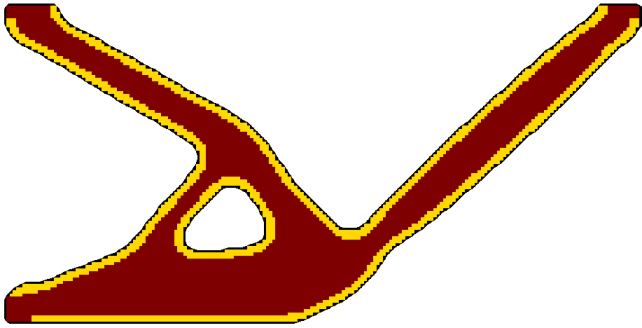}		
	&   \includegraphics[width=0.9\linewidth]{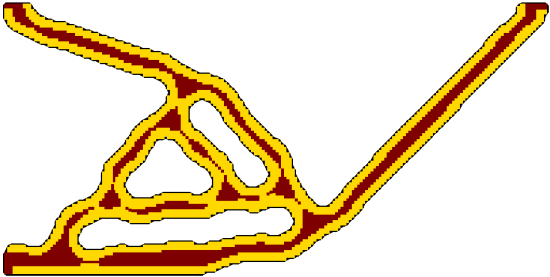}	
	&   \includegraphics[width=0.9\linewidth]{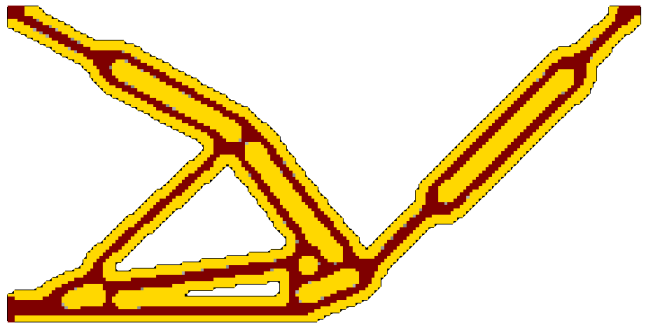}
	\\
	\cmidrule(r){1-4} 
	\multirow{3}{*}[5.0ex]{\rotatebox{90}{\small{perimeter path}}}
    &   \includegraphics[width=0.9\linewidth]{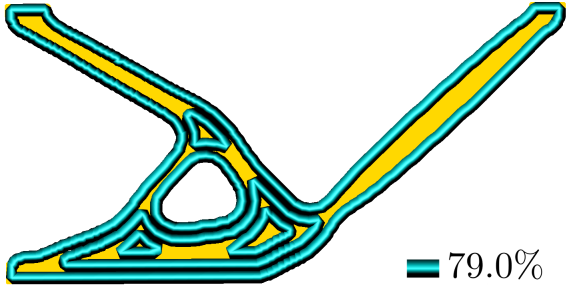}		
	&   \includegraphics[width=0.9\linewidth]{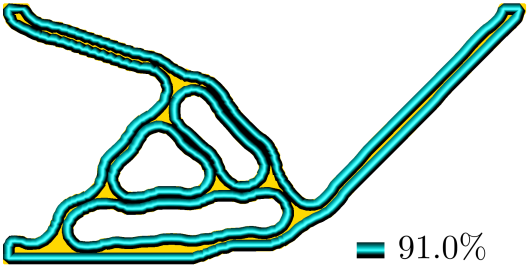}	
	&   \includegraphics[width=0.9\linewidth]{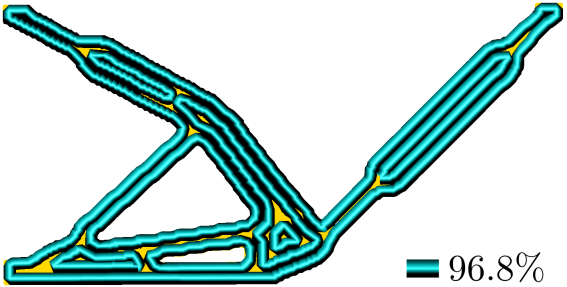}
	\\
	\bottomrule	
\end{tabular}
\label{TAB:Other_Benchmarks}	
\end{table}

Here, once again, the reference, the maximum size-constrained, and the AM-constrained problems are solved. The force inverter design solutions for a volume contraint of $30\%$ are sumarized in Table \ref{TAB:Other_Benchmarks}, along with the eroded fields and the deposition paths (note that due to symmetry, only half the domain is modeled). These results exhibit the same pattern observed in previous examples, namely, the maximum size-constrained design achieves better printability than the reference design by improving conformability to the width of the material beads. However, the maximum size constraint could introduce a significant performance reduction, as it provides low design flexibility by imposing a uniform size of solid members throughout the design. On the other hand, the AM constrained design besides providing a good printability, offers more design flexibility since it allows the creation of voluminous zones that improve performance. In addition, as a by-product, Method 2 yields an eroded design that can be interpreted as the deposition path.

\section{Conclusions and Perspectives}  \label{sec:6}

This work aims to propose topology-optimized designs suitable to Large-Scale Additive Manufacturing (LSAM) processes. The design limitation considered here is the small resolution of LSAM processes resulting from depositing large volumes of material to reduce printing time. The process resolution, which is typically defined by the size of the deposition nozzle, is introduced in the topology optimization problem through a geometric control of structural features. Specifically, two methods based on minimum and maximum size control have been proposed. The first method imposes the minimum and maximum size of the solid members in proportion to the size of the deposition nozzle. This results in designs with structural features of uniform width, which matches a discrete number of material beads arranged in a parallel fashion. The method has been evaluated in 2D and 3D design spaces, showing the capability to produce optimized designs with improved manufacturability. Interestingly, when a fixed building orientation is chosen for a 3D design problem, the method generates structures composed of walls, which are suited for fabrication by most LSAM processes. However, this method can lead to substantial reduction of performance due to the splitting of structural members as enforced by the maximum size restriction. The second method instead dilates a maximum-size constrained design which, due to a specific choice of length scales, resembles a structural skeleton or the deposition path. The dilation operation of this skeleton produces a design tailored to the deposition nozzle size, allowing for an efficient arrangement of the material beads within the design domain. Numerical results indicate that this second approach also produces designs that conform to the specified resolution, and, in addition, it comes with a lesser performance reduction compared to the first method. This is caused by the fact that the second method allows for thicker members to be created using parallel beads, in regions where this is beneficial.

\begin{figure}
	\centering
    \includegraphics[width=0.7\linewidth]{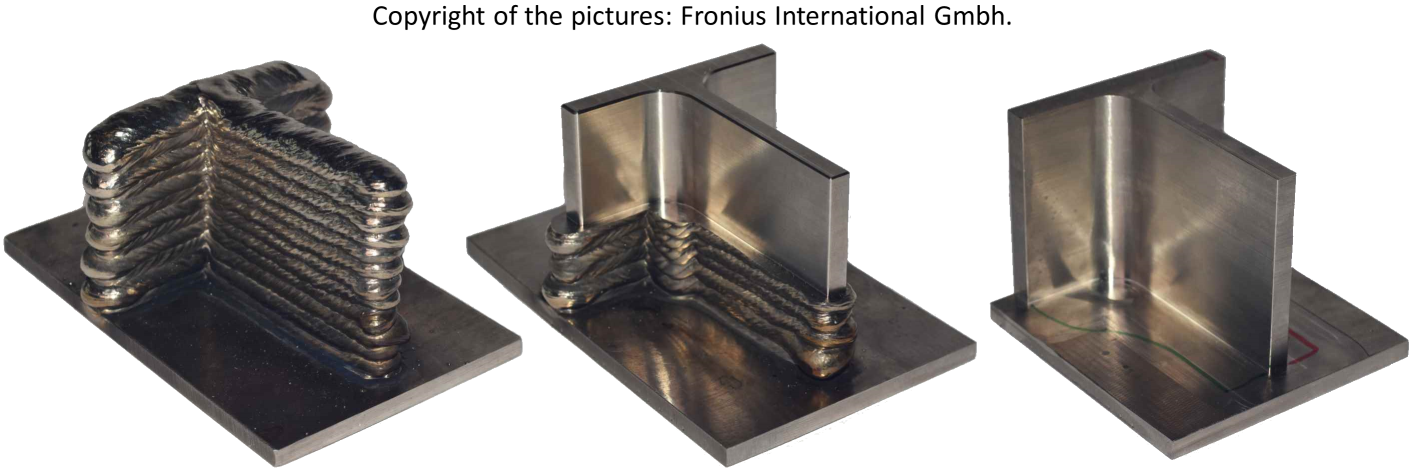}
    \vspace{-2mm}\\
    \subfigure[$\bm{\bar{\rho}}^\mathrm{dil}$]{ \hspace{20mm}
		\label{FIG:Machining_a}
		}
    \hspace{0mm} ~
	\subfigure[Machining]{ \hspace{30mm}
		\label{FIG:Machining_b}
		}
	\hspace{4mm} ~
	\subfigure[$\bm{\bar{\rho}}^\mathrm{int}$]{ \hspace{20mm}
		\label{FIG:Machining_c}
		}
	\\ 
	\caption{Post machining of a WAAM component. (a) The component obtained after the WAAM process. (b) The component undergoing machining. (c) The final component. Pictures are courtesy of Fronius International Gmbh (www.fronius.com).}	
	\label{FIG:Machining}			
\end{figure}

In light of the numerous design guidelines that must be taken into account when designing for LSAM, it is reasonable to think of combining the proposed topology optimization methods with others existing in the literature. For example, the 3D designs obtained for a fixed printing direction (Table \ref{TAB:Results_3D_2}) could be difficult to manufacture with certain LSAM processes due to the presence of extreme overhang angles. When the use of additional support structures is not considered or feasible, the introduction of overhang angle constraints \citep{Langelaar2016} in the topology optimization problem could be considered. Also, depending on the application, it may be necessary to post-machine a component manufactured by WAAM for instance, to improve surface finish and mechanical performance. As illustrated in Fig.~\ref{FIG:Machining}, the post-machining operation slims the WAAM component, similar to an erosion process of the robust formulation. Therefore, it is likely feasible to consider post-processing in the adopted formulation. Certainly, the access of the machining tool should also be considered, as in the topology optimization methods presented in \citep{Li2016,Langelaar2019,Gaynor2020}. These ideas and the combination of the proposed methods with other Topology Optimization methods are subjects of future research.

\section*{Acknowledgements}

The corresponding author acknowledges the research project FAFil (\textit{Fabrication Additive laser par d\' {e}p\^{o}t de Fil
}), funded by INTERREG V A Grande Région and the European Regional Development Fund (ERDF). Computational resources have been provided by the \textit{Consortium des \'{E}quipements de Calcul Intesif} (C\'{E}CI), funded by the Scientific Research Fund of Belgium (F.R.S.-FNRS)

\bibliographystyle{tfcad}
\bibliography{interactcadsample}

\end{document}